\documentclass[fleqn,usenatbib]{mnras}

\usepackage{newtxtext,newtxmath}

\usepackage[T1]{fontenc}

\DeclareRobustCommand{\VAN}[3]{#2}
\let\VANthebibliography\thebibliography
\def\thebibliography{\DeclareRobustCommand{\VAN}[3]{##3}\VANthebibliography}

\usepackage{textcomp, gensymb}
\usepackage{kotex}
\usepackage{comment}
\usepackage{amsmath}
\usepackage{xspace}
\usepackage{graphicx}
\usepackage{rotating}
\usepackage{color}

\newcommand{\lya}        {Ly$\alpha$\xspace}
\newcommand{\hi}         {H~{\sc i}\xspace}

\newcommand{\NHI}        {\relax\ifmmode{N_{\rm HI}\xspace} \else {$N_{\rm HI}$}\expandafter\xspace\fi}
\newcommand{\cl}      {\ifmmode{N_{\rm HI,cl}\xspace}\else{$N_{\rm HI,cl}\,$\xspace}\fi}

\newcommand{\kms}        {\ifmmode{\rm \,km\,s^{-1}}\else\,km\,s$^{-1}$\xspace\fi}
\newcommand{\unitNHI}    {\ifmmode{\rm \,cm^{-2}}\else\,cm$^{-2}$\xspace\fi}  
\newcommand{\vexp}       {\relax\ifmmode {v_{\rm exp}} \else {$v_{\rm exp}$}\expandafter\xspace\fi}
\newcommand{\sigsrc}     {\relax\ifmmode {\sigma_{\rm src}} \else {$\sigma_{\rm src}$}\expandafter\xspace\fi}
\newcommand{\dvpeak}     {$\Delta V_{\rm peak}$\xspace}

\newcommand{\anglelos}        {\relax\ifmmode{\theta_{\rm los}\xspace} \else {$\theta_{\rm los}$}\expandafter\xspace\fi}
\newcommand{\vrot}        {\relax\ifmmode{V_{\rm rot}\xspace} \else {$V_{\rm rot}$}\expandafter\xspace\fi}

\newcommand{\rin}{$R_{\mathrm{i}}$}
\newcommand{\rout}{$R_\mathrm{o}$}
\definecolor{myblue}{rgb}{0.13, 0.13, 0.8}

\title[Ly$\alpha$ RT in a rotating medium]{
The Impact of Circumgalactic Rotation on Ly$\alpha$ Radiative Transfer
}

\author[H. Kym et al.]{
Hee-Gyeong Kym,$^{1,2}$
Seok-Jun Chang,$^{3}$\thanks{Corresponding Author, E-mail: sjchang@mpa-garching.mpg.de}
Max Gronke,$^{4}$
and Kwang-Il Seon$^{1,2}$
\\
$^{1}$Korea Astronomy and Space Science Institute, 776 Daedeokdae-ro, Yuseong-gu, Daejeon 34055, Korea\\
$^{2}$Department of Astronomy and Space Science, University of Science and Technology, 217, Gajeong-ro, Yuseong-gu, Daejeon 34113, Korea\\
$^{3}$Max-Planck-Institut f\"{u}r Astrophysik, Karl-Schwarzschild-Stra$\beta$e 1, 85748 Garching b. M\"{u}nchen, Germany \\ 
$^{4}$Astronomisches Rechen-Institut, Zentrum f\"{u}r Astronomie, Universit\"{a}t Heidelberg, M\"{o}nchhofstra$\beta$e 12-14, 69120 Heidelberg, Germany}

\date{Accepted XXX. Received YYY; in original form ZZZ}
\pubyear{\the\year{}}

\begin{document}
\label{firstpage}
\pagerange{\pageref{firstpage}--\pageref{lastpage}}
\maketitle

\begin{abstract}
Hydrogen Lyman-alpha (Ly$\alpha$) is a prominent emission line from the circumgalactic medium (CGM). Due to its resonant nature, Ly$\alpha$ carries imprints of the physical properties and kinematics of the cold CGM. In particular, CGM rotation can modify the Ly$\alpha$ peak separation, which is often interpreted as a tracer of H~I column density. We present 3D Monte Carlo Ly$\alpha$ radiative-transfer simulations in a CGM-like rotating medium and examine how the emergent spectra depend on rotational velocity ($V_{\rm rot}$), H~I column density ($N_{\rm HI}$), viewing angle, clumpiness, and intrinsic source width.
We find that rotation broadens the integrated spectra and increases the peak separation, with the strongest viewing-angle dependence when rotational Doppler shifts dominate over frequency diffusion. At high $N_{\rm HI}$, numerous scatterings reduce the sensitivity of integrated spectra to rotation, producing a degeneracy between $V_{\rm rot}$ and $N_{\rm HI}$. Consequently, Ly$\alpha$ peak separation alone can overestimate $N_{\rm HI}$ in a rotating medium. Spatially resolved halo spectra provide a clearer diagnostic: opposite sides of the rotating medium show systematic redshifted and blueshifted asymmetries associated with the line-of-sight velocity of the last-scattering gas. Such rotation-driven signatures can also contribute to velocity-map patterns often interpreted in terms of inflow or outflow, highlighting the need to consider rotation in spatially resolved Ly$\alpha$ observations. We further show that the main signatures persist in simple clumpy media, while the halo signatures are largely insensitive to the intrinsic source width. Our results demonstrate that spatially resolved Ly$\alpha$ observation is essential for disentangling CGM rotation from radiative-transfer effects.
\end{abstract}



\section{Introduction} \label{sec:intro}

The circumgalactic medium (CGM) is the gas in the outskirts of galaxies bounded by their dark matter halos, and it serves as a bridge between the intergalactic medium (IGM) and the interstellar medium (ISM). The CGM is a key component of galaxy evolution, acting as a reservoir of baryons---including metals---and connecting galaxies to their larger-scale environments. Understanding the physical conditions and kinematics of the CGM provides valuable insight into the baryon cycles around galaxies and the feedback processes of star formation \citep[see reviews][]{tumlinson2017, faucher-giguere2023}.  

One important aspect of CGM kinematics is rotation, which is linked to angular momentum transport, disk formation, and the exchange of gas between galaxies and the IGM \citep{stewart2017, defelippis2020}. Observational studies have searched for signatures of CGM rotation using absorption-line measurements toward background quasars and integral-field unit (IFU) spectroscopy of extended emission \citep{hodges-kluck2016, zabl2019, nateghi2024}. In particular, absorption-line studies suggest that cool CGM gas may share the sense of rotation of the galactic disk \citep{steidel2002, bouche2013, ho2017, french2020, kacprzak2010, kacprzak2025}. However, direct constraints on CGM rotation remain challenging.

Among the various tracers of the CGM, \lya emission stands out as one of the most prominent. It is the brightest hydrogen line at high redshift and has been widely observed in the form of extended \lya halos surrounding galaxies \citep{steidel2011, wisotzki2016, leclercq2017, arrigoni_battaia2019, gonzalez_lobos2023}. These halos provide spatially resolved information about the distribution and kinematics of the CGM gas. 

\lya photons undergo multiple scatterings, producing extended halos and characteristic spectral features such as double-peaked or asymmetric line profiles. The resonant nature of \lya makes its radiative transfer highly sensitive to the properties of neutral hydrogen, including column density, temperature, and velocity fields \citep{adams1972, neufeld1990,verhamme2006, zheng2002, laursen2009, dijkstra2019}.
In addition, the small-scale structure of neutral hydrogen can also affect \lya radiative transfer \citep{neufeld1991,Hansen2006,gronke2016,chang2023}, as observational studies suggest that neutral gas in the CGM may be distributed in small clouds with high covering factors and low volume filling factors \citep{rauch1999, prochask2009, lan2017}.
These complexities have motivated various Monte Carlo and analytical studies of \lya radiative transfer \citep{seon2020, almadaMonter2024}.

Previous theoretical studies of \lya radiative transfer have primarily focused on static gas or radial flows (inflow and outflow) to explain observed line asymmetries \citep{ahn2002, dijkstra2006, dijkstra2012}. However, the impact of rotation on \lya transfer has received little attention, despite its potential importance for interpreting spatially extended emission.
In the case of a static, optically thick \hi medium, \lya radiative transfer has been extensively studied and is known to produce a symmetric double-peaked profile \citep{harrington1973, neufeld1990}. The large scattering cross-section at line center causes \lya photons to experience multiple scatterings in the medium before escaping \citep{adams1972}. As \lya photons scatter through the medium, their frequencies shift, and both the peak separation and peak width increase with increasing optical depth. Thus, the peak has been used as an indicator of H I column density \NHI.

The \lya spectrum emerging from a rotating medium also exhibits a double-peaked profile, as shown in previous studies using idealized geometries \citep{zheng2002, garavito2014, remolina-gutierrez2019}. In rigidly rotating models, the emergent spectrum depends on the viewing angle, rotational velocity, and optical depth, with higher rotational velocities producing broader profiles and enhanced line-center flux \citep{garavito2014}. This framework was subsequently extended to include radial outflows while preserving the same rigid rotation model \citep{remolina-gutierrez2019}.

However, due to the degeneracies between spectra produced by a static, optically thick medium and those emerging from a rotating medium, interpretation of the resulting spectra is challenging. Furthermore, previous studies have primarily considered idealized solid-body rotation, while the effects of more realistic rotation curves on \lya spectra remain unclear.

In this study, we present 3D Monte Carlo radiative transfer simulations of \lya in the CGM, in which the gas exhibits rigid-body rotation in the inner part of the galaxy and co-rotation with the disk at a constant speed in the outer region, consistent with typical galactic rotation curves. Our goal is to examine how rotation affects \lya observables, including spectral line profiles and spatial morphology, and to identify distinct signatures that can be tested with current and future observations. 

The paper is organized as follows. In Section~\ref{sec:RT}, we briefly introduce \lya scattering and describe our radiative transfer code. In Section~\ref{sec:results}, we present the results of our simulation in a rotating medium. In Section~\ref{sec:discussion}, we discuss topics relevant to the present results.

\section{Radiative Transfer Simulation} \label{sec:RT}

In this section, we introduce our radiative transfer simulation and describe its setup based on the 3D radiative transfer code \texttt{RT-scat} \citep{chang2024}.

\subsection{Scattering Geometry} \label{sec:geometry}

\begin{figure*}
    \includegraphics[width=\textwidth]{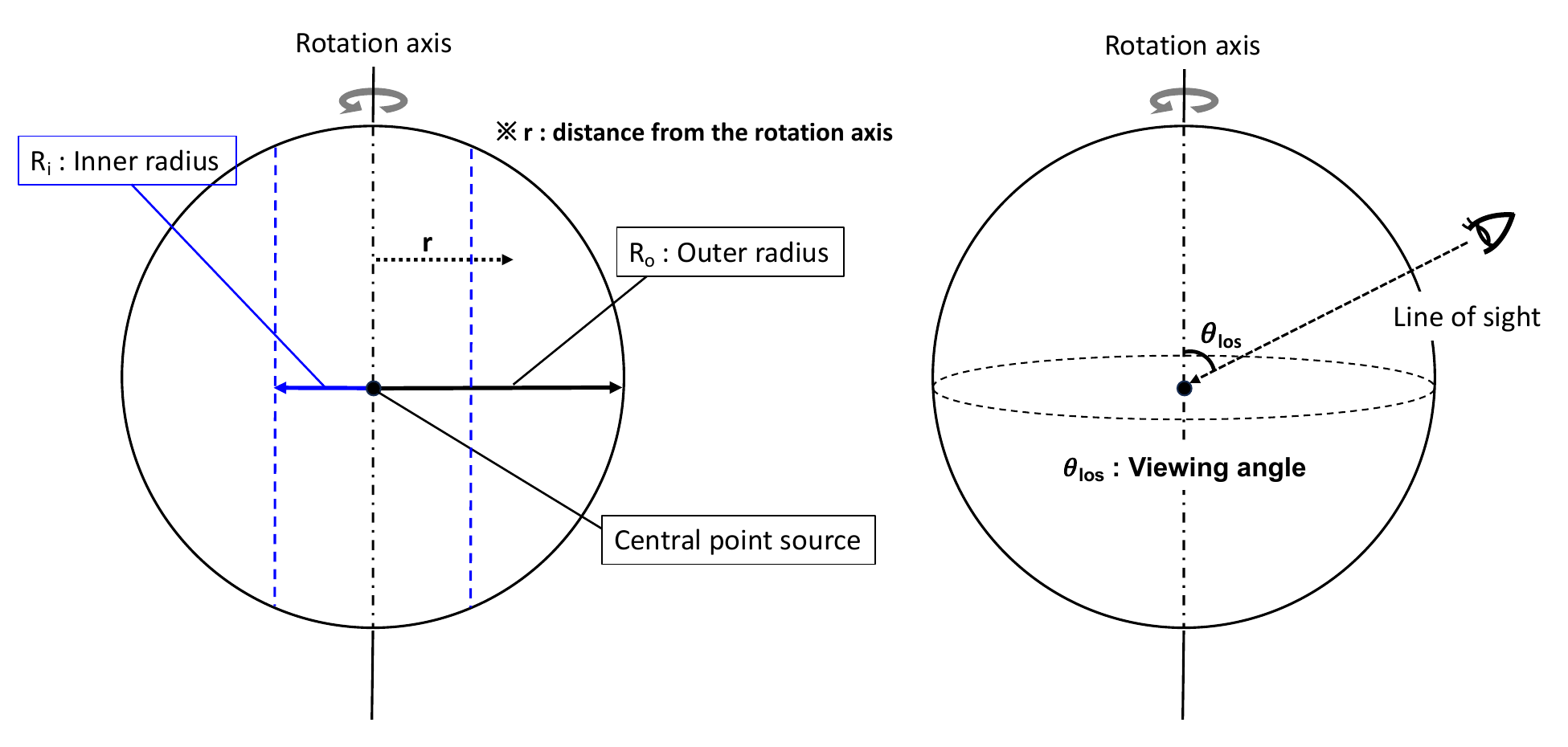}
    \caption{
    Geometry of the rotating medium. The left panel represents a projected view of the medium, viewing perpendicular to the rotation axis. The blue dashed lines indicate how the inner radius separates the region $r < $ \rin\  and \rin\  $ <r< $ \rout. The right panel shows how the viewing angle (\anglelos) is defined. The viewing angle is the angle between the rotation axis and the observer's line of sight.
    }
    \label{fig:geometry}
\end{figure*}

We assume a spherical medium with a point source at its center. We test two types of density distributions for the medium: smooth and clumpy. In the smooth medium case, \hi atoms are homogeneously distributed throughout the sphere.
The clumpy medium case represents a multiphase scenario in which spherical clumps are randomly distributed within the sphere, and the space between them is empty. The clump radius is defined to be 1/1000 of the medium's radius. We use the covering factor $f_c$ as a measure of how densely clumps populate the medium. It is defined as the mean number of clumps intersected by a radial line of sight from the center to the boundary of the sphere. Note that for such a clumpy medium, the geometry relevant to the radiative transfer is primarily characterized by $f_c$, with little dependence on the individual clump sizes or their size distribution \citep{Hansen2006,chang2023}.

Figure~\ref{fig:geometry} shows the geometry of the rotating medium. The medium rotates about the $z$-axis according to the right-hand rule. The medium is characterized by an inner radius (\rin) and an outer radius (\rout). The inner radius is set to $R_{i} = 0.1R_{o}$.
The rotation curve of the medium adopted in this paper is inspired by studies of Mg~{\sc ii} absorption lines outside the galactic plane \citep{ho2017}, and is defined as 

\begin{equation}
V(r)=
\begin{cases}
V_{\rm rot}\left(\dfrac{r}{R_{\rm i}}\right), & r\leq R_{\rm i},\\[4pt]
V_{\rm rot}\ , & R_{\rm i} < r \leq R_{o},
\end{cases}
\end{equation}

The rotational velocity increases proportionally with the distance from the rotation axis in the inner region ($r < $ \rin), and remains constant ($=\vrot$) in the outer region ($r > $ \rin). Here, $r$ is the distance from the rotation axis. The viewing angle (\anglelos) is defined as the angle between the rotation axis and the observer's line of sight. The edge-on and face-on views correspond to $\anglelos = 90\degree$ and $0\degree$, respectively.

Photon packets are emitted isotropically from the central point source.
We consider a monochromatic source and Gaussian source spectra centered at the \lya wavelength with widths \sigsrc.
We examine rotation effects with respect to the following parameters shown in Table~\ref{tab:parameters}: rotational velocity (\vrot), column density (\NHI), the line width of the source spectrum (\sigsrc) in units of km s$^{-1}$, and viewing angle (\anglelos).
The number of photon packets was 10$^6$ for \NHI = 10$^{18-19}$ cm$^{-2}$, and 10$^5$ for \NHI = 10$^{20-21}$ cm$^{-2}$. We used fewer photons in the higher column density cases to reduce the computational time.

\begin{table}
    \centering
    \caption{Parameters used in this study. Here, \vrot\ denotes the rotation speed in the outer region of the model medium, \NHI\ is the column density measured along the radial direction from the center, \sigsrc\ is the source line width, and \anglelos\ is the viewing angle.}
    \label{tab:parameters}
    
    \begin{tabular}{@{\hspace{1.2em}}c@{\hspace{9em}}c@{\hspace{1.2em}}}
        \hline
        \textbf{Parameter} & \textbf{Values} \\
        \hline
        $\vrot$ (\kms) & 0,\; 25,\; 50,\; $\cdots$,\; 450 \\
        $\anglelos$ (\textdegree) & 0,\; 15,\; 30,\; 45,\; 60,\; 75,\; 90 \\
        $\NHI$ (cm$^{-2}$) & $10^{18}$,\; $10^{19}$,\; $10^{20}$,\; $10^{21}$ \\
        $f_{c}$ & 1,\; 2,\; 5,\; 10,\; 20,\; 50,\; 100 \\
        \sigsrc (\kms) & 0\textsuperscript{*},\; 100,\; 200,\; 300 \\
        \hline
        \noalign{\vskip -1pt}
        \multicolumn{2}{r}{\footnotesize \textsuperscript{*} Monochromatic source.}
    \end{tabular}
\end{table}

\subsection{Monte-Carlo simulation} \label{sec:monte-carlo}

\texttt{RT-scat} is a 3D grid-based Monte-Carlo code that uses a Cartesian coordinate system. Each grid cell is assigned a rotation velocity and temperature according to the input parameters. The center of the medium and the location of the photon source are defined as the origin of the system. The rotation axis is aligned with the $z$-axis, so the velocity component along this axis is set to zero. Only motions in the $xy$ plane are included in the velocity field.
The medium is assumed to have a constant temperature of $T=10^4$ K, corresponding to a thermal velocity of 12.8 km s$^{-1}$.
Dust is not included in the present study.

Our code follows five steps,

\begin{enumerate}
\item Geometry and Kinematics - We define the geometry and assign kinematic properties based on the input parameters, including the rotational velocity, density, and temperature.

\item Photon Generation — Photon packets are emitted isotropically from the central point source. Depending on the setting, the initial frequency distribution is either monochromatic or Gaussian.
For the Gaussian spectrum, the standard deviation ($\sigma_{\rm src}$) is varied over the range 100–300 km s$^{-1}$, as shown in Table~\ref{tab:parameters}.

\item Propagation and Scattering — For each photon, the path length to the next scattering event is sampled from an exponential distribution. At the scattering site, the photon's frequency and direction are updated based on the medium's local velocity, the scattering atom's velocity, and the scattering phase function (see \citealt{seon2020}).

\item Iteration — Step (3) is repeated until the photon escapes the medium.

\item Repetition — The procedure is repeated for the total number of photons.
\end{enumerate}

We compute the emergent spectra by recording the propagation direction and frequency of every photon packet that leaves the computational domain. The viewing-angle dependence is then obtained by binning the escaped photons according to the angle between their propagation direction and the rotation axis. We obtain the output \lya\ spectra for viewing angles of $\anglelos = 0, 15, \cdots, 90\degree$ in increments of 15$\degree$, as shown in Table \ref{tab:parameters}. In addition to the spectrum integrated over the entire medium, we also separately construct spectra for the halo and core regions, defined by projected radii of $r_p > 0.1R_o$ and $r_p < 0.1R_o$, respectively.

\section{Results} \label{sec:results}

In this section, we present simulated results of \lya radiative transfer in the rotating medium.
In Section~\ref{sec:monochromatic}, we first examine the case of a monochromatic \lya source to establish a fundamental understanding of how rotation, column density, and viewing angle shape the emergent spectra.
In Section~\ref{sec:clumpy}, we extend our analysis to a clumpy medium and investigate how gas inhomogeneity modifies the \lya spectral features by rotation.
In Section~\ref{sec:vemit}, we explore the dependence on the intrinsic source spectrum by considering Gaussian emission profiles with different velocity widths.

\subsection{Monochromatic Light}  \label{sec:monochromatic}

\begin{figure*}
    \includegraphics[width=0.49\linewidth]{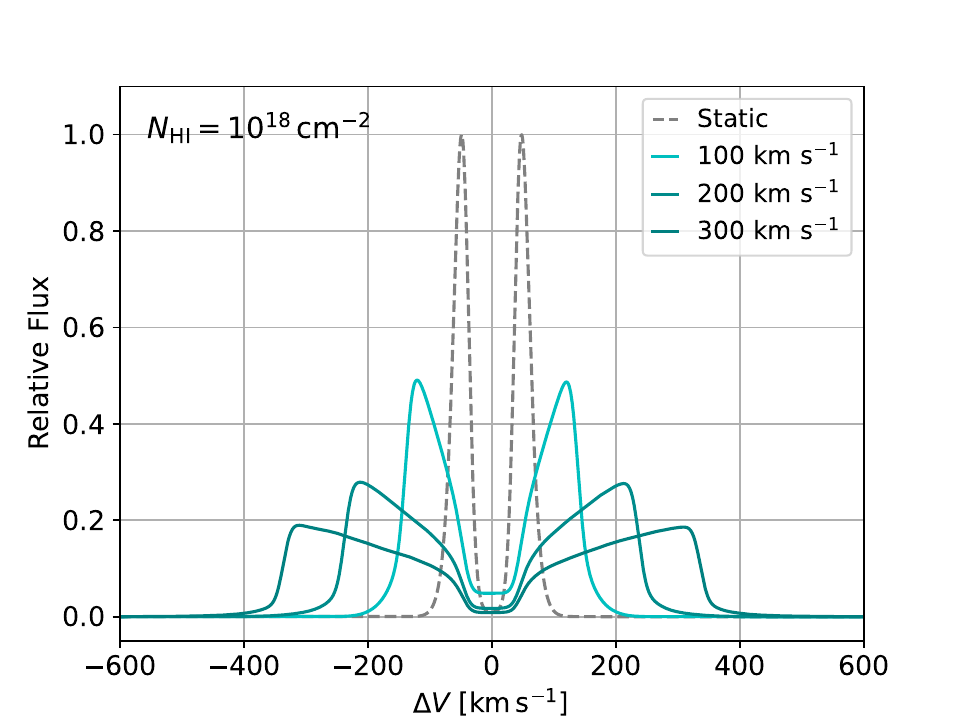}
    \includegraphics[width=0.49\linewidth]{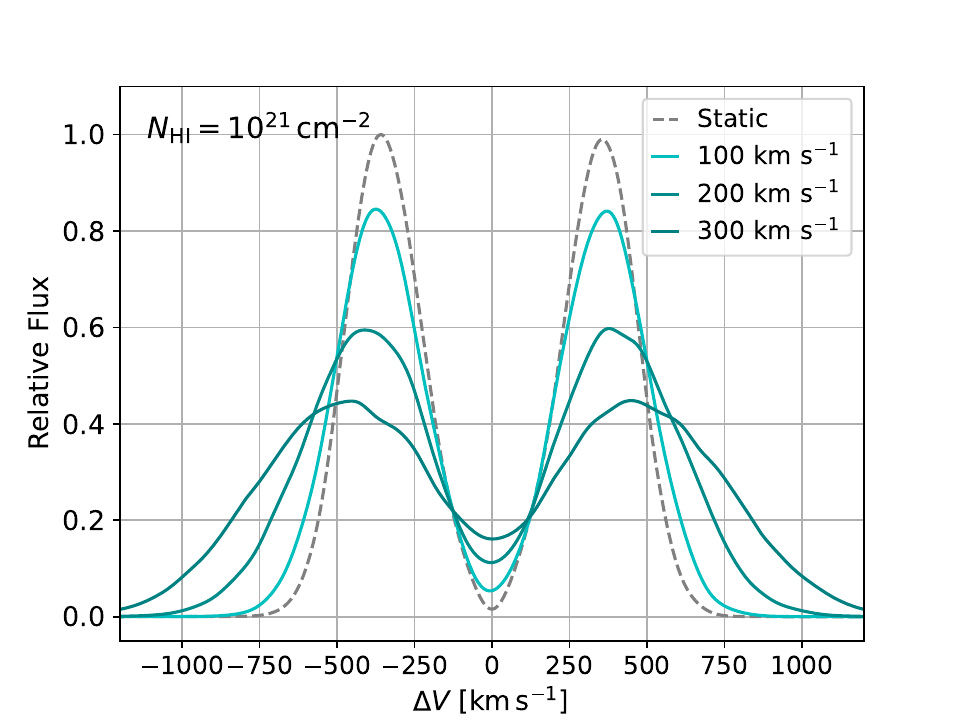}
    \caption{
    Integrated \lya spectra of various rotational velocities \vrot at \anglelos = 90\textdegree. \NHI is fixed at 10$^{18}$ and 10$^{21}$ \unitNHI in the left and right panels, respectively. Solid colored lines correspond to rotating cases with different \vrot, while dashed lines show the static case (\vrot = 0~\kms). In the low \NHI regime, the separation of the double peaks increases clearly with increasing \vrot, indicating that rotational Doppler shifts dominate the spectral formation.
    }
    \label{fig:vrot_effect}
\end{figure*}

The \lya spectrum emerging from a rotating medium generally exhibits a double-peaked profile.
Figure~\ref{fig:vrot_effect} shows integrated \lya spectra for various rotating velocities \vrot = 0 (static), 100, 200, and 300~\kms, observed perpendicular to the rotation axis (i.e., \anglelos = 90\degree). As \vrot increases, the spectra become broader, and the separation between the red and blue peaks increases.
This trend arises because the rotation of the \hi gas induces additional Doppler shifts during scattering, effectively broadening the frequency distribution of escaping photons.
However, the separation of the double peaks has traditionally been interpreted as a tracer of the \hi column density \citep{verhamme2006}. Disentangling the effects of rotation from those of numerous scatterings in the static medium, therefore, requires a clear physical understanding of \lya radiative transfer in a rotating medium.

\begin{figure}
    \centering
    \includegraphics[width=\linewidth]{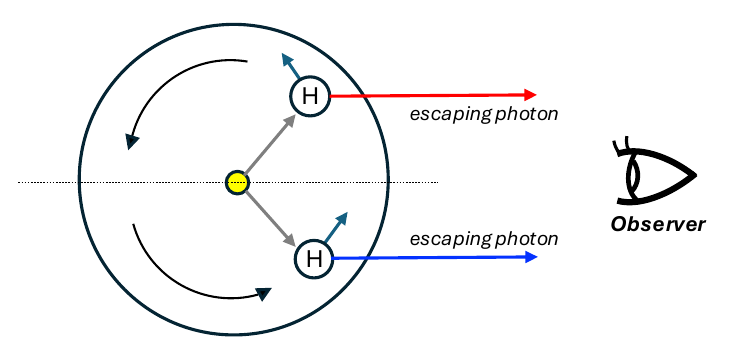}
    \caption{
    Schematic illustration of \lya photon scattering in a rotating medium with low column density ($\NHI \lesssim 10^{18} \unitNHI$).
    The illustration shows a top view of the system, where the rotation axis is perpendicular to the observer’s line of sight.
    Depending on the location of the last scattering, photons escaping from gas receding from the observer are redshifted ($\Delta V > 0 \kms$), while those escaping from approaching gas are blueshifted ($\Delta V < 0 \kms$).
    }
    \label{fig:schematic_scattering}
\end{figure}

Figure~\ref{fig:schematic_scattering} schematically illustrates the formation of the double-peaked \lya profile in a low-\NHI rotating medium, where the optical depth is sufficiently low that \lya photons undergo only a few scatterings before escaping.
For \lya photons that have just been emitted from the source and have not yet experienced any scattering, the rotating medium effectively appears static because its rotational motion is perpendicular to the photon propagation direction. As a result, the emergent wavelength is primarily determined by the location and direction of the last scattering event.
Therefore, the scattered direction determines the observed frequency. In the edge-on view (\anglelos = 90\degree), gas rotating toward and away from the line of sight produces the blue- and red-shifted components, respectively.
This causes additional broadening of the \lya profile beyond that produced by scattering-induced frequency diffusion.
Because this Doppler shift depends on the projection of the rotational velocity along the line of sight, the rotational imprint on the spectrum is expected to be weakest in the face-on view (\anglelos = 0\degree).

However, the dominant parameter of \lya radiative transfer is a \hi column density \NHI (i.e., optical depth), which determines the number of scattering.
Since each \lya scattering induces a frequency shift on the scale of \hi thermal motion, leading to a symmetric double-peaked profile due to numerous scatterings in a static medium, the impact of rotating motion on \lya profile depends on \hi column density, the key parameter controlling the number of scatterings. 
This behavior is evident in Figure~\ref{fig:vrot_effect}.
In the left panel, corresponding to low \NHI ($10^{18}\,\unitNHI$), the peak separation increases markedly with \vrot.
In the right panel, corresponding to high \NHI ($10^{21}\,\unitNHI$), the spectra are dominated by frequency diffusion, and the dependence on \vrot becomes significantly weaker.

\subsubsection{Dependence on rotational velocity \vrot}
\label{sec:vrot}

\begin{figure}
    \includegraphics[width=1\linewidth]{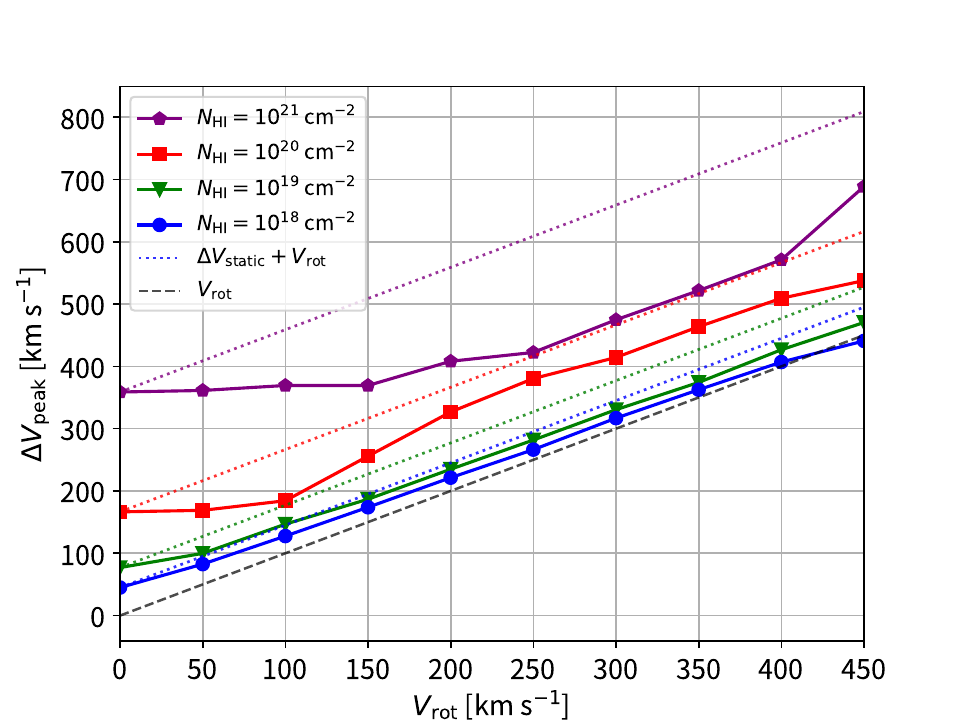}
    \caption{
    Peak positions (\dvpeak) with various rotational velocity (\vrot) for four different column densities $\NHI = 10^{18-21} \unitNHI$ in viewing angle \anglelos = 90\degree. The black dashed line shows $\Delta V_{\rm peak} = \vrot$ (i.e., $y=x$), and the dotted lines indicate \dvpeak $= \Delta V_{\rm static} + \vrot$ for each \NHI, where $\Delta V_{\rm static}$ is the peak position in the static case in the corresponding \NHI. These lines are added as a reference of linearity.}
    \label{fig:vpeak_vs_vrot}
\end{figure}

\begin{figure*}
    \centering
    \includegraphics[width=0.49\linewidth]{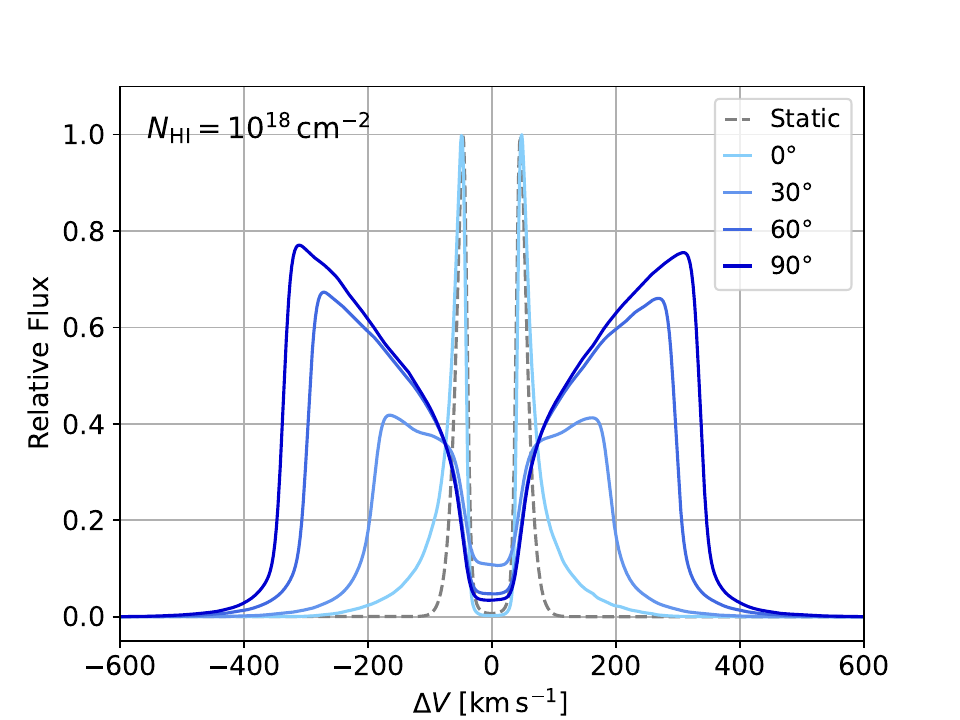}
    \includegraphics[width=0.49\linewidth]{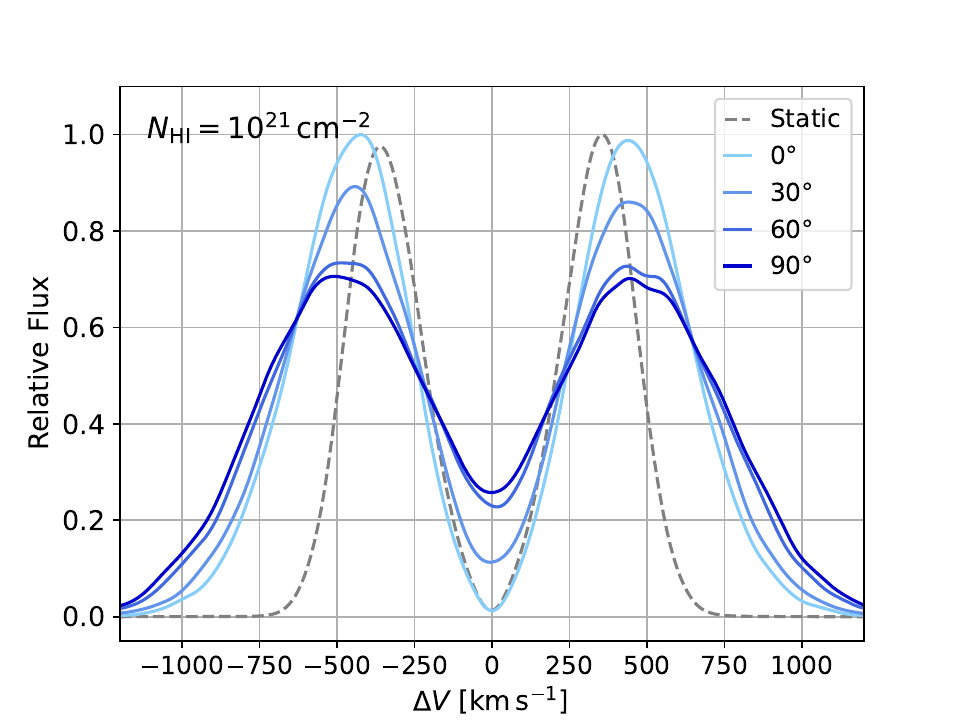}
    \caption{Integrated \lya spectra at different viewing angles for a rotating medium with $\vrot = 300\kms$. The \hi column density of the medium is $\NHI$ = $10^{18} \rm cm^{-2}$(left) and $\NHI$ = $10^{21} \rm cm^{-2}$(right) in each panel. The peak separation increases with the viewing angle in the low \NHI regime, while the dependence becomes much weaker at high \NHI.
    }
    \label{fig:viewing_angle_effect}
\end{figure*}

To quantify the dependence of the peak separation on \vrot, we measure the peak positions (\dvpeak) of the \lya spectrum in the edge-on (\anglelos = 90\degree) direction in Figure~\ref{fig:vpeak_vs_vrot}.
Because the emergent spectra are approximately symmetric, we define \dvpeak as the average of the absolute velocities of the red and blue peaks relative to line center. For an approximately symmetric double-peaked profile, the peak separation is therefore \(2\Delta V_{\rm peak}\).
Figure~\ref{fig:vpeak_vs_vrot} shows \dvpeak as a function of \vrot for different values of \NHI, demonstrating that \dvpeak increases with \vrot\ regardless of \NHI.

However, this increasing trend is not monotonic, as the dominant mechanism determining the peak position changes with \NHI.
In the low-\NHI regime ($\NHI \le 10^{19}$~\unitNHI), \dvpeak shows a nearly linear correlation with \vrot,
since Doppler shifts associated with the rotational motion dominate the formation of the double-peaked profile.
The peak position of the static medium $\Delta V_{\rm static}$ characterized by \NHI is small compared to the Doppler shift introduced by rotation. Consequently, the emergent peak position is primarily determined by the projected rotational velocity, producing an approximately linear increase of \dvpeak with \vrot.
As a result, the measured peak shifts closely follow \dvpeak$~\simeq \Delta V_{\rm static} + \vrot$, as shown by the dotted lines in Figure~\ref{fig:vpeak_vs_vrot}.

As \NHI gets higher, peak shift by frequency diffusion (i.e, $\Delta V_{\rm static}$)
becomes comparable to, or larger than \vrot, and the contribution of rotation to \dvpeak becomes weaker. 
This transition from the Doppler-dominated regime and the diffusion-dominated regime occurs when the rotational velocity becomes comparable to the static peak position ($\vrot\sim \Delta V_{\rm static}$). 
Figure~\ref{fig:vpeak_vs_vrot} shows that the \dvpeak--\vrot relation at $\NHI = 10^{21} \unitNHI$ flattens for $\vrot < \Delta V_{\rm static}$, whereas the dependence of \dvpeak on \vrot becomes stronger for $\vrot > \Delta V_{\rm static}$.
As a result, increasing \vrot no longer produces a proportional increase in \dvpeak in the high \NHI case.

We will discuss the physical origin of this trend in more detail using spatially resolved spectra in Section~\ref{sec:resolved_spectra}

\subsubsection{Dependence on viewing angle \anglelos} \label{sec:viewing_angle}

\begin{figure}
    \centering
    \includegraphics[width=1\linewidth]{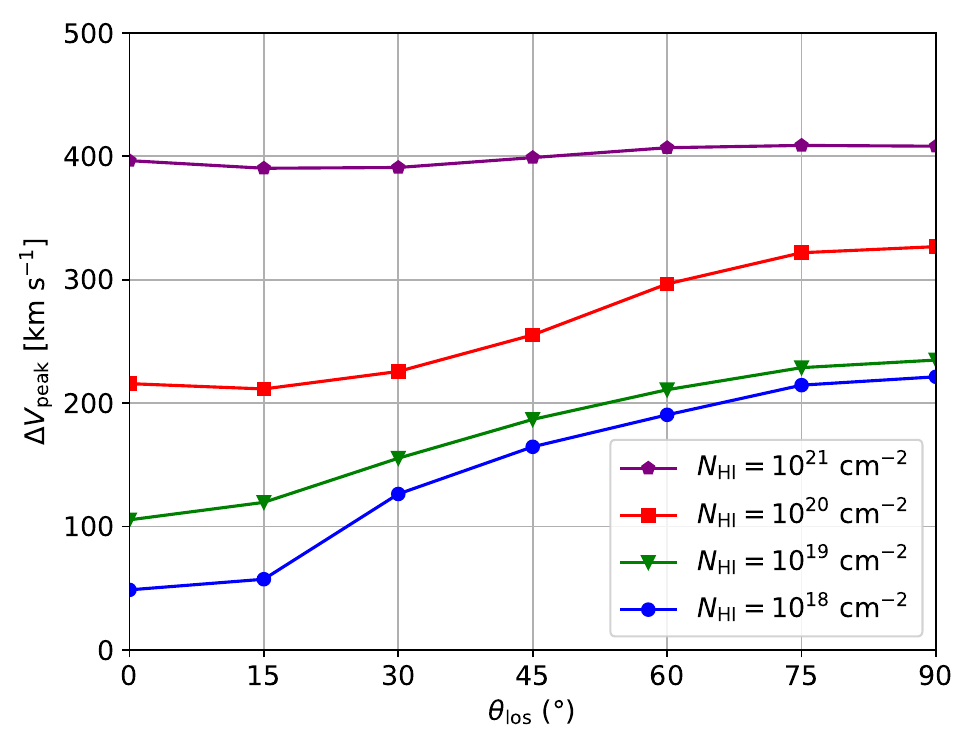}
    \caption{Peak positions \dvpeak of the integrated spectrum as a function of viewing angle $\theta_{\rm los}$ and \hi column density. The rotational velocity \vrot is fixed at 200 \kms. Each color represents a different column density, $\NHI$ = $10^{18}$ (blue), $10^{19}$ (green), $10^{20}$ (red), $10^{21} \unitNHI$ (purple). As $\NHI$ increases, \dvpeak increases with increasing $\theta_{\rm los}$ when $\NHI \le 10^{20} \rm cm^{-2}$. 
    At $\NHI = 10^{21} \rm cm^{-2}$, the dependence of \dvpeak on $\theta_{\rm los}$ is almost negligible.}
    \label{vpeak_vs_viewing_angle}
\end{figure}

In the previous subsection, we focused on the dependence of the \lya spectrum on \vrot. When rotational motion dominates the formation of the \lya spectrum, a strong dependence on the line of sight is expected.
Figure~\ref{fig:viewing_angle_effect} shows the \lya spectra for four viewing angles: $\anglelos = 0\degree$ (face-on), $30\degree$, $60\degree$, and $90\degree$ (edge-on).
In the left panel, at \NHI = $10^{18} \unitNHI$, the spectrum at $\anglelos = 0\degree$ is similar to that of a static medium because the line of sight is perpendicular to the rotating motion. As the viewing angle increases toward $\anglelos = 90\degree$, the peak separation also increases, reflecting the increasing contribution of rotational Doppler shifts when the velocity field is projected along the line of sight.

However, in the right panel, at \NHI = $10^{21} \unitNHI$, the dependence of the peak separation on \anglelos becomes weak.
Although photons experience Doppler shifts induced by rotation, the medium remains optically thick because of the high \NHI, and photons continue to interact with \hi gas. As a result, repeated scatterings redistribute the imprint of rotation throughout the medium, including the face-on direction ($\anglelos = 0\degree$). Consequently, even face-on spectra ($\anglelos = 0\degree$) exhibit a larger peak separation than the corresponding static case.

This behavior is summarized in Figure~\ref{vpeak_vs_viewing_angle}, which shows the peak shift \dvpeak as a function of \anglelos for different values of \NHI. Although \dvpeak generally increases with increasing \anglelos, this trend becomes less pronounced as \NHI increases. In particular, at the highest \NHI, \dvpeak appears to be nearly independent of \anglelos.
However, this independence applies only when $\vrot \lesssim \Delta V_{\rm static}$. As \vrot approaches and exceeds $\Delta V_{\rm static}$, the viewing-angle dependence gradually re-emerges.

\subsubsection{Comparison with a static medium}
\label{sec:compare_static}

\begin{figure}
    \includegraphics[width=1\linewidth]{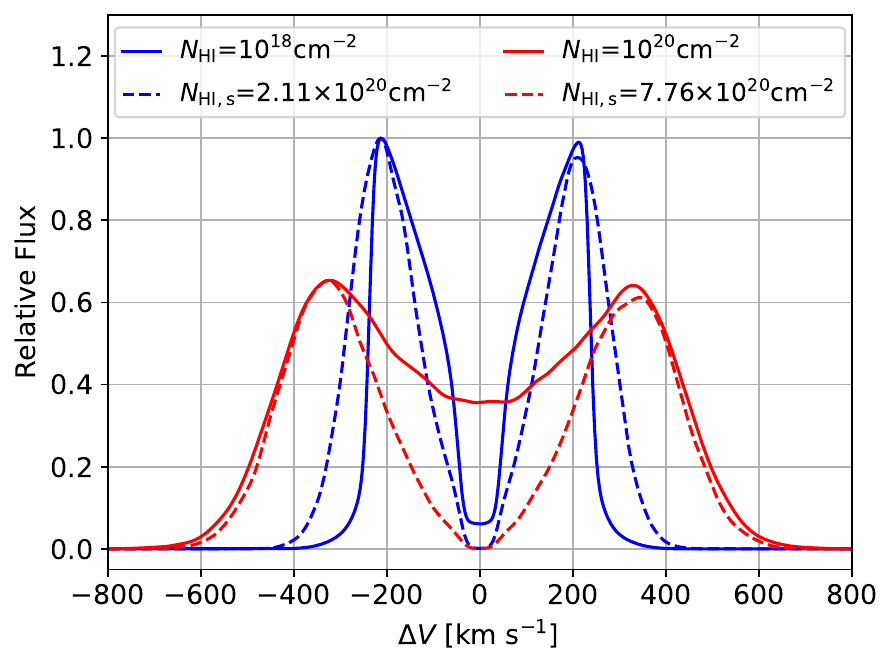}
    \caption{Spectra for rotating media with low and high H~{\sc i} column densities, together with static cases that reproduce the same peak position.
    The solid lines represent spectra of rotating media at \NHI = $10^{18} \unitNHI$(blue) and $10^{20} \unitNHI$(red) at a viewing angle 90\textdegree. The rotational velocity is fixed at \vrot = 200 \kms. The dashed lines show static cases with the same peak position as the rotating cases at each \NHI.
    }
    \label{fig:peaksep_overplot}
\end{figure}

In the previous sections, we characterized the dependence of the \lya spectra on the rotational velocity \vrot and the viewing angle \anglelos using the peak position \dvpeak.
Although \dvpeak alone cannot uniquely distinguish the effects of \vrot and \NHI, the overall spectral shapes differ between rotating and static media. 

Figure~\ref{fig:peaksep_overplot} compares the spectra of the rotating models with two column densities, $\NHI = 10^{18}$ and $10^{20} \unitNHI$, at \vrot = 200 \kms and \anglelos = 90\degree, with those of static models that have the same \dvpeak.
To reproduce the same $\Delta V_{\rm peak}$ as the rotating case with $\NHI = 10^{18}$~\unitNHI, the static medium requires a column density larger by a factor of $\sim 200$. For $\NHI = 10^{20}$~\unitNHI, the corresponding factor is reduced to $\sim 8$.
This difference reflects the distinct physical mechanisms responsible for the formation of the double-peaked profile: frequency diffusion through multiple scatterings in static media and Doppler shifts due to bulk rotation.

In the lower \NHI case ($10^{18} \unitNHI$), the spectrum of the rotating case is slightly broader than that of the counterpart of the static case in Figure~\ref{fig:peaksep_overplot}, even though the static counterpart has a much higher \NHI. 
However, at $\NHI = 10^{20}$~\unitNHI, even a smaller difference in \NHI between the rotating and static cases leads to a pronounced difference in flux near the line center, while the widths and spectral shapes outside the double peaks remain similar. The line-center flux has been used as a proxy for LyC escape \citep{gazagnes2020,naidu2022}. We will discuss this in Section~\ref{sec:discussion_line_center_flux}.

These differences in the spectral profile demonstrate that \dvpeak alone cannot unambiguously distinguish the effects of rotation from those of optical depth. Spatially resolved spectra may offer a more direct probe of rotational Doppler shifts, since the two sides of the rotation axis retain information about the line-of-sight rotational velocity of the scattering medium, as illustrated in Figure~\ref{fig:schematic_scattering}.

\subsubsection{Spatially resolved spectra}
\label{sec:resolved_spectra}

\begin{figure*}
    \centering
    \includegraphics[width=0.49\linewidth]{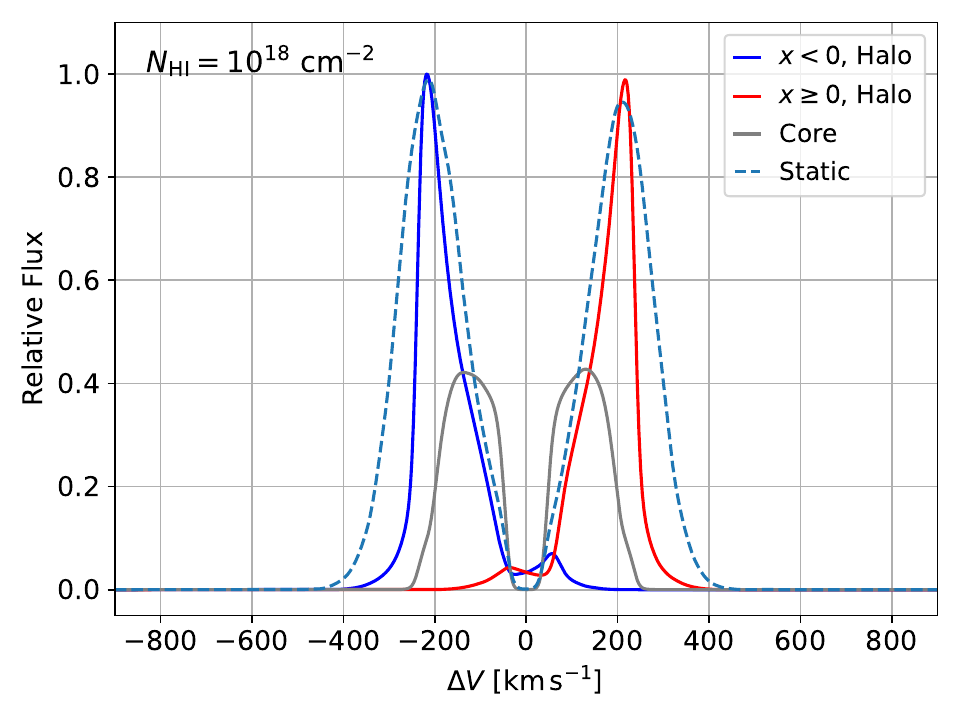}
    \includegraphics[width=0.49\linewidth]{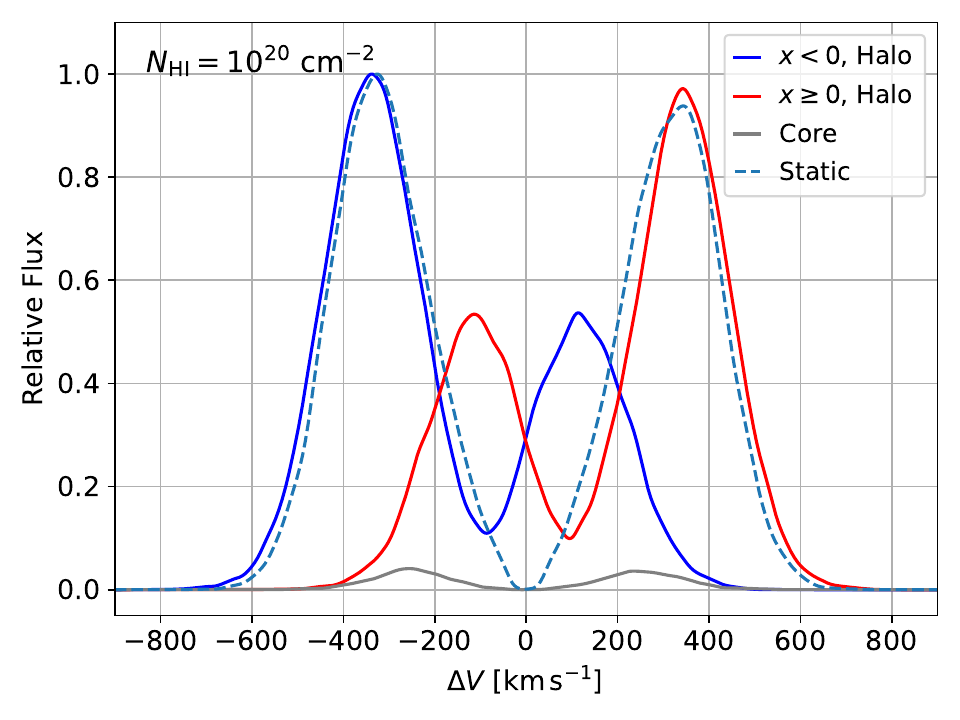}
    \caption{Spatially resolved spectra for rotating media with \vrot = 200\kms and \anglelos = 90\textdegree\, for two column densities: \NHI = $10^{18}\unitNHI$(left) and \NHI = $10^{20}\unitNHI$(right). Halo spectra are plotted as blue and red solid lines, and the core spectrum is shown as a gray solid line. For comparison, a halo spectrum of the static case with the same peak position as the integrated spectra of the rotating case is shown as a dashed line.}
    \label{fig:resolved_spec}
\end{figure*}

To understand the physical origin of the integrated spectra presented above, we now examine spatially resolved spectra of the rotating medium.
Figure~\ref{fig:resolved_spec} shows spatially resolved \lya spectra from rotating and static media.
The core spectra are obtained from photons with projected radius $r_p < 0.1 R_o$, while the halo spectra are obtained at $r_p > 0.1 R_o$.
In the rotating medium, the halo spectra differ systematically between opposite sides of the source: photons escaping from gas moving toward the observer are blueshifted, whereas those escaping from gas moving away are redshifted (see Figure~\ref{fig:schematic_scattering}).

In the left panel of Figure~\ref{fig:resolved_spec}, the halo spectra for $\NHI = 10^{18} \unitNHI$ show single, blue- and red-shifted peaks. In this regime, rotational Doppler shifts dominate the formation of the emergent spectra, and the peak offsets scale directly with the rotational velocity,
consistent with the behavior shown in Figure~\ref{fig:vpeak_vs_vrot}.

However, in the right panel of Figure~\ref{fig:resolved_spec}, corresponding to the high \NHI regime ($\NHI = 10^{20} \unitNHI$), the halo spectra display clear double-peaked profiles with asymmetric peak strengths.
In this regime, as the Doppler shift from the rotating motion is insufficient to make photons escape, the photons undergo additional scattering. As a result, frequency diffusion produces the characteristic double-peaked structure, while rotation shifts the spectra on opposite sides of the halo in opposite directions.

The superposition leaves the integrated peak positions nearly unchanged when $\Delta V_{\rm static} \gg \vrot$. As \vrot increases, the outermost peaks increasingly determine the maxima of the integrated profile, leading to the re-emergence of the dependence of \dvpeak on \vrot. \\

In summary,  \lya radiative transfer in a rotating medium with a monochromatic source produces double-peaked spectra whose formation is governed by the relative contributions of rotational Doppler shifts and frequency diffusion through multiple scatterings.
At low \hi column density, rotation strongly influences both \dvpeak and its dependence on viewing angle, with \dvpeak scaling closely with \vrot.
At high \hi column density, frequency diffusion becomes increasingly important, reducing the sensitivity of the integrated spectra to both \vrot and \anglelos.
This behavior occurs when the rotational velocity is smaller than the peak position of the static medium (\vrot $\lesssim \Delta V_{\rm static}$). As \vrot approaches and exceeds $\Delta V_{\rm static}$, \dvpeak gradually recovers its dependence on \vrot\ .
Although these competing effects introduce strong degeneracies in spatially integrated spectra, spatially resolved spectra retain clear kinematic signatures of rotation through oppositely shifted peak structures. We will discuss the spatially resolved case in Sections~\ref{sec:discussion_spatially_resolved} and \ref{sec:discussion_velocity_map}.

\subsection{Dependence on Clumpiness}\label{sec:clumpy}

\begin{figure*}
    \centering
    \includegraphics[width=0.49\linewidth]{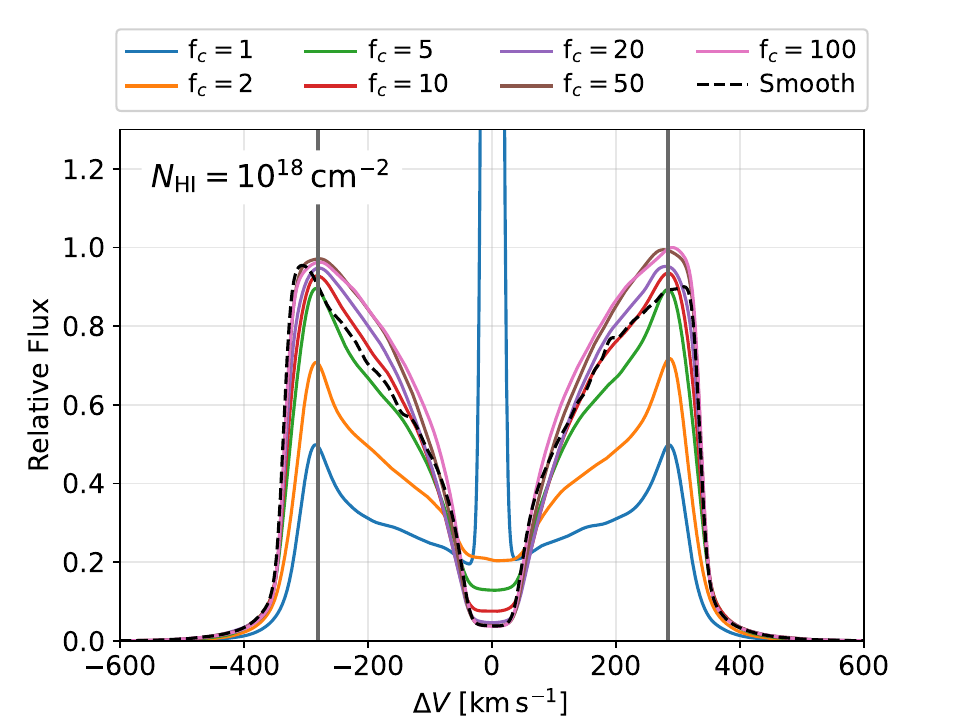}
    \includegraphics[width=0.49\linewidth]{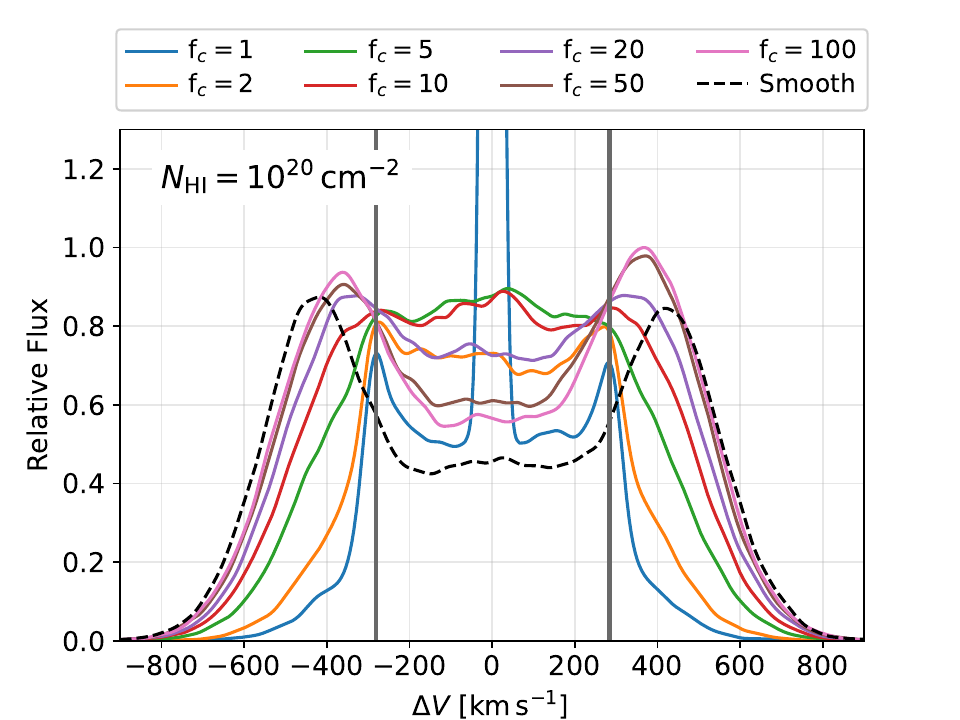}
    \caption{\lya profiles of rotating clumpy media for two H~{\sc i} column densities, \NHI$=10^{18}$ (left) and $10^{20}$ cm$^{-2}$ (right). The rotational velocity and viewing angle are fixed at $\vrot=300\kms$, and \anglelos = 90\degree. The covering factor ranges from $f_{c}=$1 to 100. The smooth medium case is shown by a black dashed line. For reference, the vertical gray lines mark the peak positions of the rotating clumpy medium cases with $\NHI=10^{18}\unitNHI$.}
    \label{fig:clumpy_fc}
\end{figure*}

Previous studies of \lya radiative transfer in multiphase media have considered optically thick neutral clumps, where photons can undergo repeated scatterings at clump surfaces \citep{neufeld1991,Hansen2006}. In such clumpy configurations, the emergent \lya spectrum approaches that of a smooth medium as the covering factor ($f_c$) increases \citep{gronke2016,gronke2017a,chang2023}.

Here, we adopt a simplified setup in which spherical clouds of neutral hydrogen with a size of $r_{\rm cl}=10^{-3}$\rout\ are embedded in a fully ionized surrounding medium. The clumps are uniformly distributed within the spherical halo and do not overlap. We explore covering factors in the range $f_c = 1$–$100$ and average total H~{\sc i} column densities along a sightline from the halo center to the surface in the range $N_{\rm HI,,tot}=10^{18}$–$10^{21}$ cm$^{-2}$. In this setup, the total column density is related to that of individual clumps by $N_{\rm HI,,tot} = (4/3) f_c N_{\rm HI,,cl}$, where the factor of $4/3$ is a geometrical factor \citep{gronke2016,gronke2017a}. The number of clumps and the volume filling factor are given by $N_{\rm cl}=(4/3)f_c(R_o/r_{\rm cl})^2$ and $f_{\rm v}=N_{\rm cl}(r_{\rm cl}/R_o)^3$, respectively.
We use this setup to test whether the rotational signatures identified in the smooth model survive in an inhomogeneous medium. Since the effect of rotation is maximized in the edge-on direction, we focus on the $\anglelos = 90\degree$ case.

Figure~\ref{fig:clumpy_fc} shows \lya\ emerging spectra from rotating, clumpy media for various covering factors $f_c$, observed at $\anglelos = 90^\circ$, for two H~{\sc i} column densities, $\NHI = 10^{18}$ and $10^{20}\, \unitNHI$.
As $f_c$ increases, the spectral profiles of the clumpy medium gradually converge to those of the smooth medium, consistent with previous studies of \lya transfer in multiphase media described above. 

As discussed in Section~\ref{sec:monochromatic}, the dominant mechanism shaping the \lya line profile depends on \NHI: rotational Doppler shifts primarily determine the peak structure at low \NHI, whereas frequency diffusion through multiple scatterings dominates at high \NHI.
The same behavior is clearly seen in the clumpy medium results.

In the left panel of Figure~\ref{fig:clumpy_fc}, corresponding to $\NHI = 10^{18}\,\unitNHI$, the peak separation in clumpy media is nearly identical to that in the smooth medium, even for small $f_c$.
In this low \NHI regime, \lya photons typically experience only a small number of scatterings before escape, and a single interaction with a rotating clump (i.e., at low $f_c$) is sufficient to imprint the characteristic rotational Doppler shift. As a result, the peak separation is largely insensitive to $f_c$.

\begin{figure*}
    \centering
    \includegraphics[width=1\linewidth]{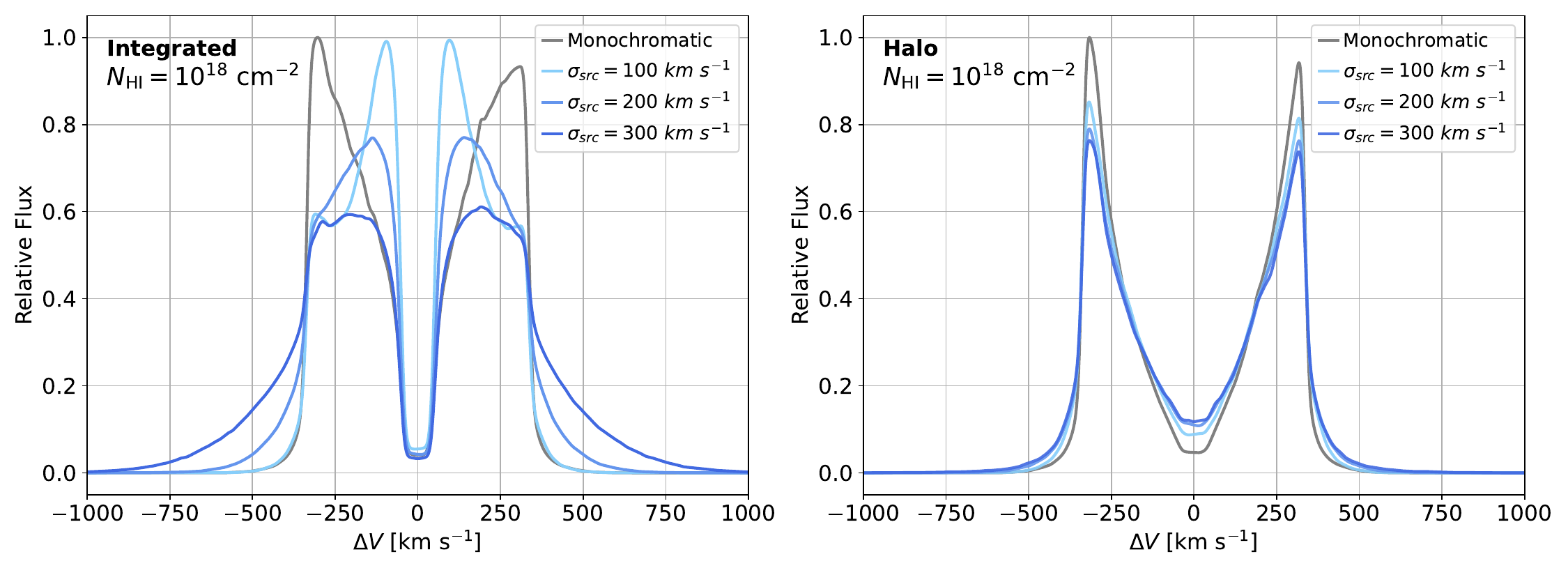}
    \caption{Integrated (left panel) and halo (right panel) spectra for different source widths, monochromatic and \sigsrc = $100, 200,$ and $300\kms$. The column density, viewing angle, and rotational velocity of the medium are fixed at \NHI=$10^{18}$\unitNHI, \anglelos=90\textdegree, \vrot=300\kms, respectively. The integrated spectra are normalized by their intensities, so that the areas under the spectra are identical. The halo spectra are normalized to have identical peak intensities. Different \sigsrc s are shown in different colors.}
    \label{fig:diffinput}
\end{figure*}

\begin{figure*}
    \centering
    \includegraphics[width=1\linewidth]{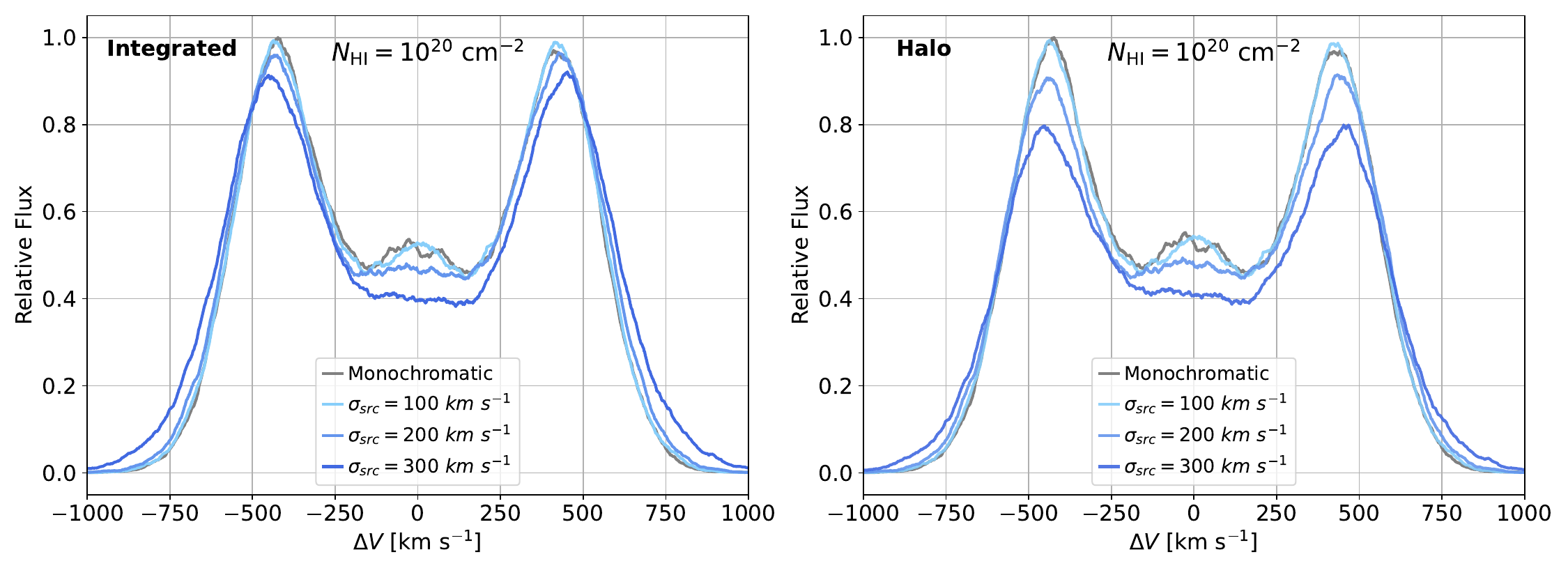}
    \caption{Integrated (left panel) and halo (right panel) spectra for different source widths, monochromatic and \sigsrc = $100, 200,$ and $300\kms$. The column density, viewing angle, and rotational velocity of the medium are fixed at \NHI = $10^{20}$\unitNHI, \anglelos = 90\textdegree, \vrot = 300\kms, respectively. The integrated spectra are normalized by their intensities, so that the areas under the spectra are identical, and the same normalization factors are applied to the halo spectra.}
    \label{fig:diffinput_highNHI}
\end{figure*}

However, the right panel of Figure~\ref{fig:clumpy_fc} shows that, for $\NHI = 10^{20}\, \unitNHI$, the peak separation depends strongly on $f_c$.
When $f_c < 10$, the spectra exhibit significantly narrower double peaks than those of the smooth medium or clumpy models with a large covering factor ($f_c > 10$).
These narrower peaks resemble those seen in the low \NHI case.
At low $f_c$ case, photons interact with only a limited number of clumps and undergo few scatterings within each clump. Consequently, frequency diffusion due to repeated scatterings is inefficient and cannot produce the broad peaks characteristic of the smooth medium with high \NHI.

In summary, at low \NHI, the \lya peak separation is largely independent of $f_c$, as rotational Doppler shifts dominate the spectral formation.
At high \NHI, however, the peak structure becomes sensitive to $f_c$. For low $f_c$, the reduced number of interactions with clumps suppresses frequency diffusion, leading to spectra similar to those in lower \NHI-cases, where Doppler shift remains dominant despite the large total \NHI.

\subsection{Dependence on the Intrinsic Source Spectrum}
\label{sec:vemit}

\begin{figure*}
    \centering
    \includegraphics[width=0.48\linewidth]{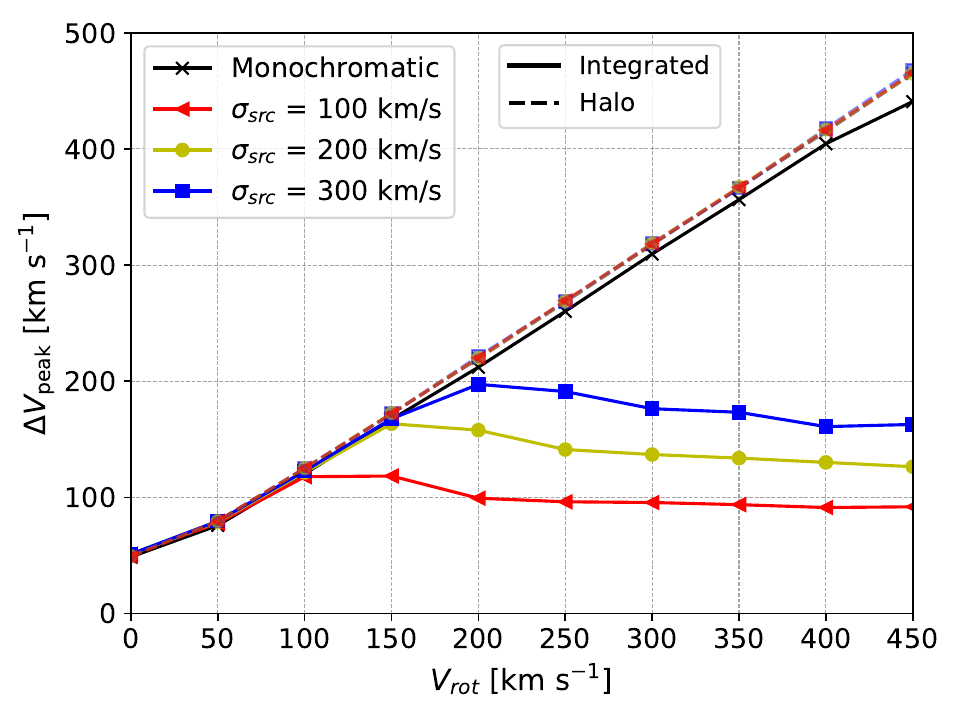}
    \includegraphics[width=0.48\linewidth]{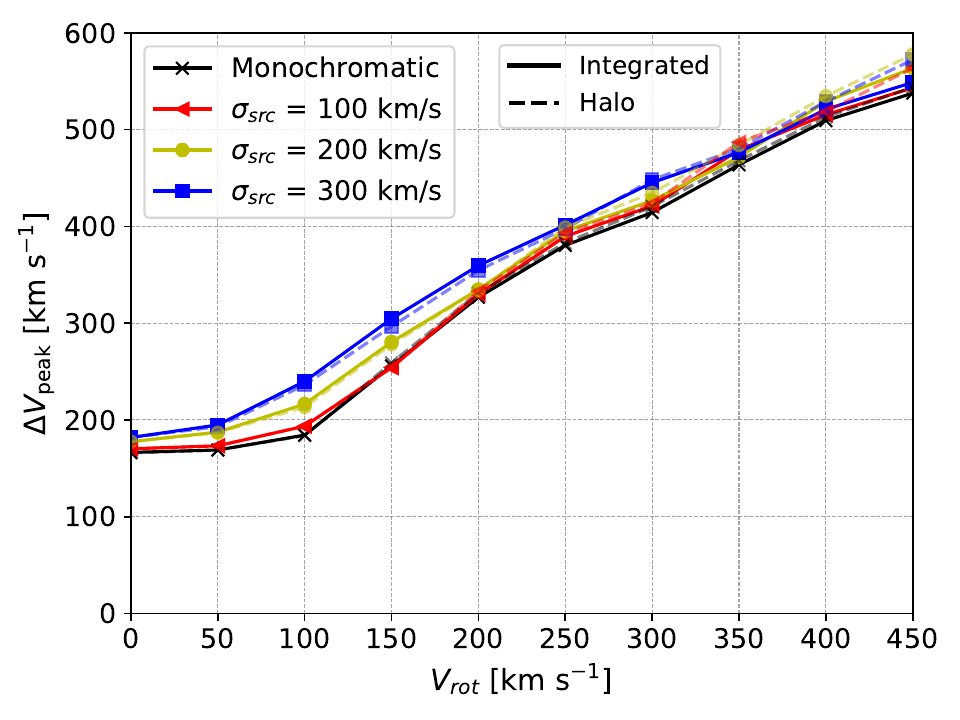}
    \caption{Peak positions of the integrated and halo spectra as a function of \vrot for four source spectra. The column density and viewing angle are fixed at \NHI = $10^{18}$\unitNHI(left), $10^{20}$\unitNHI(right), and \anglelos = 90\textdegree, respectively. Solid and dashed lines indicate the peak positions of the integrated and halo spectra, respectively.}
    \label{fig:vemit_vrot}
\end{figure*}

In Section~\ref{sec:monochromatic}, we adopted a monochromatic source at the line center to study the fundamental behavior of \lya transfer in a rotating medium.
In realistic systems, however, the intrinsic \lya emission is not monochromatic and is expected to have a finite width.
Because the \lya scattering cross section follows a Voigt profile, the propagation of \lya photons depends sensitively on the frequency distribution of the incident radiation.
To investigate this effect, we consider Gaussian source spectra with different velocity widths, $\sigma_{\rm src}$.

Figures~\ref{fig:diffinput} and \ref{fig:diffinput_highNHI} show the integrated and halo \lya spectra for a monochromatic source and Gaussian source spectra with $\sigma_{\rm src} = $ 100, 200, and 300~\kms. The two figures correspond to $\NHI = 10^{18}$ and $10^{20}\, \unitNHI$, respectively.
The halo spectra are constructed from photons collected at projected radii $r_p > 0.1 R_o$.
Regardless of \NHI, the halo spectra are nearly identical for all $\sigsrc$. By contrast, the integrated spectra exhibit a strong dependence on the source line width in the low-\NHI case ($10^{18}\, \unitNHI$), whereas the high-\NHI case ($10^{20}\, \unitNHI$) behaves much more like the halo spectra.

In the left panel of Figure~\ref{fig:diffinput}, for the low-\NHI, the Gaussian source cases ($\sigsrc = 100-300 \kms$) show double peaks that form closer to the line center than in the monochromatic case, i.e., with a smaller peak separation.
For photons emitted from the central source, the rotating medium effectively behaves as a static medium because the rotational motion is perpendicular to the radial propagation direction.
As a result, scattering primarily affects photons near the line center, where the cross section is highest, whereas photons in the wings of a broad source spectrum can escape directly without undergoing resonant scattering.
The integrated spectra, therefore, consist of a superposition of directly escaping wing photons and scattered photons originating near the line center.
This contribution from unscattered photons leads to a reduced separation of the double peaks in the Gaussian source cases.

However, the right panel of Figure~\ref{fig:diffinput} shows that the halo spectra are largely insensitive to $\sigma_{\rm src}$.
Because the halo spectra consist only of scattered photons, their shapes are determined by radiative transfer within the rotating medium rather than by the intrinsic emission profile.
To examine this behavior further, Figure~\ref{fig:vemit_vrot} presents the peak position, \dvpeak, as a function of the rotational velocity, \vrot, for both the integrated and halo spectra.
For all Gaussian source cases, the \dvpeak values of the halo spectra closely follow those of the monochromatic case.
In contrast, \dvpeak of the integrated spectra becomes nearly independent of \vrot for $\vrot \gtrsim 150 \kms$, where directly escaping photons dominate the spectral structure.

Figure~\ref{fig:diffinput_highNHI} shows that this insensitivity becomes even stronger at high \NHI. In that regime, both the integrated and halo spectra are dominated by photons that undergo many resonant scatterings before escape. The repeated frequency redistribution erases most of the memory of the initial Gaussian width, so the emergent peak positions and the overall spectral shapes vary only weakly with \sigsrc. This result demonstrates that the kinematics and physical properties of the rotating medium are imprinted on the halo spectrum largely independently of the intrinsic source profile.

\section{Discussion} \label{sec:discussion}

In this section, we place our results in the context of previous work on rotating \lya transfer and recent efforts to extract CGM kinematics from spatially resolved \lya observations.

\subsection{Comparison with rigid-body rotation models}
\label{sec:rigid vs flat}

\cite{garavito2014} studied \lya radiative transfer in a rotating medium modeled as a homogeneous sphere in rigid-body rotation. In this idealized configuration, radiative transfer within the sphere is equivalent to that in a static sphere, and rotation mainly affects the emergent photons through Doppler shifts at the last-scattering surface.

While \cite{garavito2014} focused on an ISM-scale rotating medium, we model gas on CGM scales and a broken rotation curve motivated by halo observations, with a rising inner part and a flat outer part. Our model preserves several characteristics reported for rigidly rotating media, including broader peaks and enhanced line-center flux with increasing rotational velocity and \hi column density. However, because our CGM-like model explicitly includes radiative transfer through an extended halo, it also produces additional spectral features from photons that continue to scatter in the halo after leaving the inner, rigidly rotating region.

Our CGM-like rotating model exhibits shifts in the peak positions. These shifts depend on both the rotational velocity and the viewing angle, as shown in Sections~\ref{sec:vrot} and \ref{sec:viewing_angle}. The rotational effects are most pronounced for edge-on view ($\anglelos = 90\degree$), where \dvpeak increases monotonically with rotational velocity in the low \NHI regime.

In the low \NHI regime ($\NHI \leq 10^{19}\unitNHI$), the peak velocity increases approximately in proportion to \vrot, indicating that photons escape after relatively small frequency excursions, allowing the gas bulk velocity to directly imprint on the emergent spectrum. In contrast, as \NHI increases, higher \hi column density ($\NHI \geq 10^{20}\unitNHI$) leads to a larger frequency diffusion, which broadens the double-peaked profile and increases the peak separation, producing a degeneracy with the rotational effect.
The peak separation alone cannot constrain whether the observed profile is driven mainly by rotation or by a large H I column density. Appendix~\ref{sec:appendix_nhi_degeneracy} illustrates this degeneracy by showing the static models required to match the peak separations of the rotating cases at $\anglelos = 90\degree$; higher column densities in the static models are required to reproduce the peak separations of the rotating media. This is precisely why additional diagnostics, especially spatially resolved \lya spectra, are required.

\begin{figure*}
    \centering
    \includegraphics[width=1\linewidth]{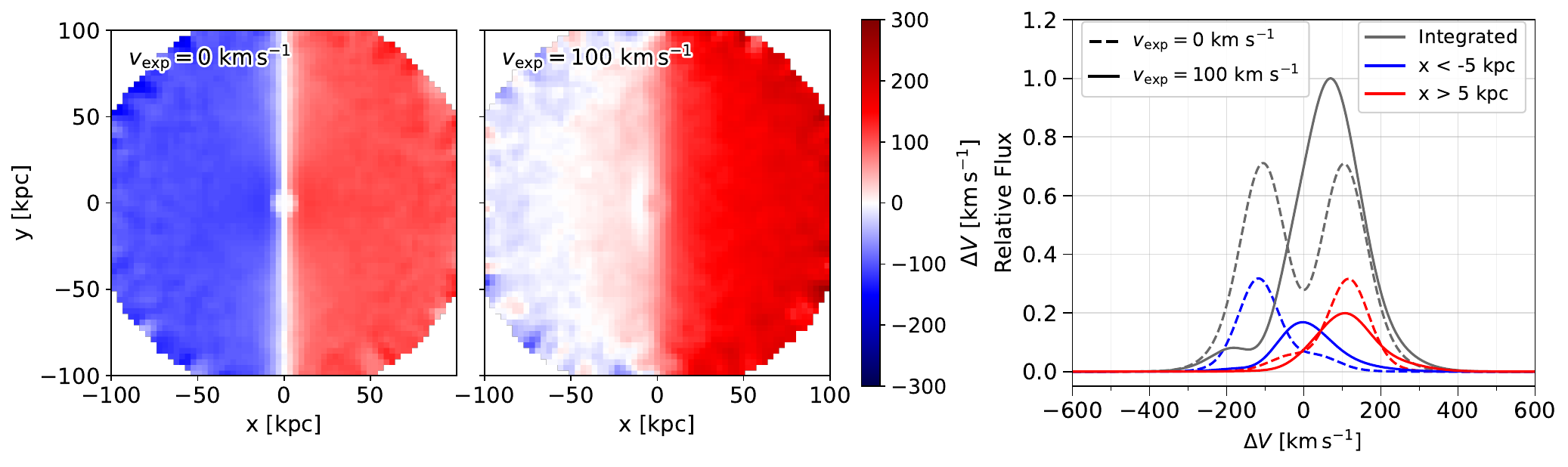}
    \caption{First-moment velocity maps and emergent spectra for models with $\NHI=10^{18}\unitNHI$, $\sigsrc=100\kms$, $\vrot=100\kms$, and $\anglelos=90\degree$. 
    The spatial (0.8″ seeing) and spectral resolutions ($R=3000$) are adopted. The left and middle panels show the velocity maps for pure rotation ($\vexp=0\kms$), and rotation plus outflow ($\vexp=100\kms$), respectively. The right panel compares the corresponding spectra. Adding the outflow component enhances the redshifted emission and weakens/mixes the blueshifted side, producing an asymmetric morphology and spectrum despite the underlying rotating geometry.}
    \label{fig:rot&outflow}
\end{figure*}

\subsection{Spatially resolved Ly$\alpha$ constraints on CGM kinematics}
\label{sec:discussion_spatially_resolved}

Early theoretical studies showed that resonant back-scattering in an expanding neutral supershell can naturally produce asymmetric \lya profiles \citep{ahn2003}. This idea was later developed into the widely used expanding shell framework for modeling \lya radiative transfer and fitting observed spectra \citep[e.g.,][]{verhamme2006,schaerer2008,verhamme2008,hashimoto2015,yang2016,yang2017,gronke2017b}. These models have been remarkably successful in reproducing a wide variety of observed \lya spectral profiles. In particular, \citet{gronke2017b} showed that the shell model can provide excellent fits to 237 MUSE-Wide \lya spectra.

However, this success is mainly limited to the spectral dimension. The standard shell model is designed to reproduce spatially integrated line profiles, but it does not self-consistently describe the observed surface-brightness profiles of extended \lya halos. This is because its simple geometry, typically a central source surrounded by a single expanding shell, lacks the radial distributions of \lya emissivity, neutral gas, dust, and velocity structure that determine the spatial diffusion of \lya photons. Consistently, \citet{gronke2017b} found no strong correlations between the best-fit shell-model parameters and the observed properties of \lya halos. Therefore, fitting the integrated spectrum alone is insufficient to constrain the physical and kinematic structure of the scattering medium in \lya halos.

To overcome this limitation, recent studies have incorporated spatial information in addition to the \lya spectrum. \citet{song2020} introduced an outflowing halo model, in which the distributions of \lya sources and H~{\sc i} plus dust are spatially extended, and simultaneously fitted both the \lya spectrum and surface-brightness profile of MUSE \lya emitters. This approach was extended by \citet{yu2025} to a larger sample of 163 MUSE \lya-emitting galaxies, demonstrating that the joint modeling of spectra and surface-brightness profiles can constrain the spatial extent and velocity structure of the scattering medium more effectively than either observable alone. Beyond azimuthally averaged halo models, \citet{li2022} analyzed spatially resolved \lya spectra of SSA22 LAB2 with radiative-transfer modeling and showed that the observed line-profile variations trace the interplay between outflowing and inflowing gas. Similarly, \citet{guo2024} used stacked MUSE data to measure the radial evolution of average \lya profiles out to $\sim 60$ kpc, showing that spatial variations in \lya line profiles encode CGM-scale gas kinematics.

First-moment maps in spatially resolved observations of \lya halos often show mixed blue and red patterns, which are commonly interpreted as signatures of inflow or outflow \citep{herenz2020, li2021, gonzalez_lobos2023}. Spatially resolved observations of giant \lya nebulae have also revealed large-scale ordered velocity gradients, which have been interpreted as signatures of rotating structures (e.g., \citealt{prescott2015,martin2016}) or inspiraling gas motions (e.g., \citealt{arrigoni_battaia2018,zhang2023sci,zhang2023apj}).

Our results show that rotating H~I gas can also produce systematic spatial variations in the Ly$\alpha$ line profile. As shown in Figure~\ref{fig:resolved_spec}, spectra extracted from opposite sides of the halo exhibit distinct red- and blueshifted asymmetries associated with the line-of-sight velocity of the scattering gas. This suggests that rotation should be considered, rather than attributing the mixed velocity patterns solely to radial motions, when interpreting spatially resolved \lya observations.

\subsection{Interpreting Ly$\alpha$ velocity maps in rotating outflows}\label{sec:discussion_velocity_map}

Figure~\ref{fig:rot&outflow} illustrates how a radial outflow modifies the Ly$\alpha$ velocity field produced by rotation.
In the left panel for the pure-rotation case, the velocity map shows a clear red-blue pattern due to the rotating gas. The approaching and receding sides are cleanly separated across the projected rotation axis, and the spectrum shows a relatively symmetric double-peaked profile. 
In the central panel for rotating outflow gas, the redshifted side becomes more dominant while the blueshifted side is weakened and mixed with velocities closer to systemic. The resulting velocity field and asymmetric spectrum could easily be interpreted as evidence for a superposition of radial motions. Our results therefore demonstrate that commonly used kinematic signatures are not unique to inflow or outflow, but can also arise from rotation alone or from a combination of rotation and radial motion.

The relative importance of rotation and radial motion in such velocity maps also depends on the H~I column density \NHI. As shown in Appendix~\ref{sec:appendix_rotating outflow}, for fixed \vrot at 100~\kms, increasing \NHI from $10^{18}$ to $10^{20}~\unitNHI$ makes the outflow signature much more prominent, with even weak outflow ($\vexp=50\, \kms$) producing a predominantly redshifted velocity field across the halo (see Figure~\ref{fig:velocity_map_high_low_NHI}). This demonstrates that multiple scattering effects can amplify the apparent imprint of radial outflow in \lya first-moment maps, further emphasizing that the observed velocity field does not directly trace the intrinsic gas motion.

\subsection{Viewing-angle effects in rotating Ly$\alpha$ halos}
\label{sec:discussion_viewing-angle}

Figure~\ref{fig:viewing_angle_effect} illustrates the dependence of the peak separation on viewing angle at two column densities. In the low-\NHI\ case ($\NHI = 10^{18}\unitNHI$), the face-on line profile ($\anglelos=0\degree$) would be practically indistinguishable, considering instrumental line broadening, from that of a static medium with the same column density: two narrow peaks at $\Delta V \approx \pm 50\kms$, set entirely by the frequency shift required for photons to escape after a few scatterings. As \anglelos\ approaches $90\degree$, however, an additional peak shift emerges, reaching $\sim\pm 300\kms$ in the edge-on view. In the high-\NHI\ case ($\NHI = 10^{21}\unitNHI$), by contrast, frequency diffusion through the optically thick medium already places the peaks near $\pm 450\kms$ in the face-on view, and varying \anglelos\ leaves the peak positions almost unchanged. The dominant orientation effect is instead a progressive broadening of the wings, together with a partial filling-in of the central trough. The peak position, \dvpeak, is therefore a sensitive tracer of rotation only in the optically thin regime, whereas at high \NHI\ the rotational signature is encoded in higher-order properties of the line, such as the FWHM and the depth of the central trough, rather than in \dvpeak\ itself. This is consistent with the trend summarized in Figure~\ref{vpeak_vs_viewing_angle}.

The viewing-angle dependence discussed in Section~\ref{sec:viewing_angle} also has practical observational consequences. Even at fixed rotational velocity, edge-on configurations maximize the line-of-sight projection of rotation and therefore produce broader spectra and larger peak separations than face-on views. This orientation effect should be taken into account when comparing \lya line widths among spatially resolved halos with different inclinations. Because the inclination of an extended \lya halo is generally not known a priori, relying only on the observed peak separation and line width may introduce a systematic bias when interpreting samples of galactic halos with different geometries. This reinforces the rotation--column-density degeneracy quantified in Appendix~\ref{sec:appendix_nhi_degeneracy} and again argues for the use of spatially resolved diagnostics to break it.

The viewing angle effects on the resonance line have also been described for other resonance lines in different geometries. \citet{seon2024} carried out Monte Carlo radiative-transfer simulations of the Mg~{\sc ii} $\lambda\lambda 2796, 2803$ doublet in static cylindrical media and found that the emergent line profile, the doublet flux ratio, and the line escape fraction depend strongly on the observer's inclination angle when the medium is flat (height-to-radius ratio $H/R_{\rm cyl} \ll 1$): photons preferentially escape along the short vertical direction in face-on views, whereas in edge-on views they must diffuse through a much larger optical path and emerge as broader spectra. The inclination dependence weakens monotonically as the geometry becomes more spherical ($H/R_{\rm cyl} \to 1$) and is further suppressed once the optical depth is high enough that repeated resonance scatterings randomize the photon directions. The saturation of the \anglelos dependence at high \NHI that we identify in Figure~\ref{fig:viewing_angle_effect} reflects the same physical principle---that resonance scattering itself washes out directional information once the medium becomes sufficiently optically thick.

\subsection{Rotation and Ly$\alpha$ peak separation}
\label{sec:discussion_peaksep}

The velocity separation between the blue and red \lya peaks is widely used as a diagnostic of the neutral hydrogen column density and optical depth of the scattering medium. In static radiative-transfer models, larger \NHI leads to stronger frequency diffusion through repeated resonant scatterings and therefore produces broader double-peaked profiles with larger peak separations (e.g., \citealt{neufeld1990,verhamme2006}). Observationally, small peak separations have also been associated with low H~{\sc i} optical depth and efficient escape of Lyman-continuum photons (e.g., \citealt{verhamme2015}). However, our results show that bulk rotational motion can also significantly increase \dvpeak, especially in the low-column-density regime where rotational Doppler shifts dominate the spectral formation (Figures~\ref{fig:vrot_effect} and \ref{fig:vpeak_vs_vrot}).

This introduces a degeneracy between rotational kinematics and optical depth in spatially integrated \lya spectra. As shown in Figures~\ref{fig:peaksep_overplot} and \ref{fig:peaksep_los&vrot}, a rotating medium with relatively low \NHI can produce peak separations similar to those of a static medium with substantially larger column density. As a result, interpreting \dvpeak solely in terms of \NHI may overestimate the true optical depth when rotational motions are present. This effect may also introduce additional scatter into empirical relations between \lya peak separation and LyC escape. Our results therefore suggest that rotational kinematics should be considered when using \lya peak separation as a tracer of CGM or ISM optical depth, and further reinforce the importance of spatially resolved \lya observations for disentangling radiative-transfer and kinematic effects.

\subsection{Line center flux at the rotating medium}
\label{sec:discussion_line_center_flux}

Previous studies have interpreted enhanced \lya line-center flux as evidence for low-opacity escape channels and have used it as a proxy for LyC escape (e.g., \citealt{verhamme2015, rivera-thorsen2017, vanzella2018, gazagnes2020, naidu2022}). More recent spatially resolved observations have suggested that the interpretation of central \lya emission can be more complex, particularly in systems with spatially varying radiative transfer conditions(e.g., \citealt{ownes2024, solhaug2025}).
Theoretical studies also have shown that \lya photons do not necessarily escape through the lowest-column density channels, complicating the interpretation of \lya observables as a tracer of LyC escape \citep{almadaMonter2024, almadaMonter2026}.

In Figure~\ref{fig:peaksep_overplot}, the enhanced flux near the line center is a distinguishable feature of the rotating medium compared with a static medium that has the same peak separation. As shown in the spatially resolved spectra (the right panel of Figure~\ref{fig:resolved_spec}), the double-peaked profiles are Doppler-shifted on either side, producing two opposite asymmetric double peaks depending on the line-of-sight rotational velocity of the medium. On the approaching side, the red peak is shifted toward the systemic velocity, while on the receding side, the blue peak is shifted toward the systemic velocity. The superposition of these oppositely shifted peaks fills the central dip, producing the enhanced line center.

In this context, our results demonstrate that gas rotation alone can also enhance the flux near the line center, even in models with high H I column densities. This suggests that gas kinematics, such as rotation, may represent an additional factor affecting this diagnostic.

\section{Conclusion}

We have presented 3D Monte Carlo \lya radiative transfer simulations in a rotating CGM-like medium. The model adopts a spherical neutral medium with a rotation curve that rises in the inner region and becomes flat in the outer halo, motivated by co-rotating CGM gas around galaxies. We explored how the emergent \lya spectra depend on rotational velocity, H~{\sc i} column density, viewing angle, clumpiness, and the intrinsic source spectrum.

Our main results are summarized as follows.
\begin{enumerate}
    \item Rotation broadens the emergent \lya spectra and increases the double-peak separation (Figures~\ref{fig:vrot_effect} and \ref{fig:vpeak_vs_vrot}). In the low-column-density regime ($\NHI\lesssim10^{19}\,{\rm cm^{-2}}$), \dvpeak increases approximately linearly with \vrot, indicating that rotational Doppler shifts dominate the spectral formation. At higher \NHI, repeated scatterings drive strong frequency diffusion and weaken the sensitivity of the integrated spectra to rotation.

    \item The rotational signature depends strongly on viewing angle when the medium is not highly optically thick (Figures~\ref{fig:viewing_angle_effect} and \ref{vpeak_vs_viewing_angle}). The peak separation is smallest in the face-on view and largest in the edge-on view, where the line-of-sight component of the rotational velocity is maximized. This viewing-angle dependence becomes weak at high \NHI because repeated scatterings isotropize the photon escape directions.

    \item Integrated spectra alone cannot uniquely distinguish rotation from optical-depth effects (Figures~\ref{fig:peaksep_overplot} and \ref{fig:peaksep_los&vrot}). A rotating low-column-density medium can produce a peak separation similar to that of a static medium with a substantially larger column density. This degeneracy implies that \dvpeak should not be interpreted as a direct tracer of \NHI without accounting for rotational kinematics.

    \item Spatially resolved \lya spectra retain clearer signatures of rotation than spatially integrated spectra (Figures~\ref{fig:resolved_spec} and \ref{fig:rot&outflow}). Halo spectra from opposite sides of the rotation axis show systematic red- and blueshifted asymmetries associated with the line-of-sight velocity of the last-scattering gas. These signatures persist even when the integrated spectra are partially degenerate with static models, demonstrating the importance of spatially resolved \lya observations.

    \item The main rotational signatures are robust against simple changes in gas structure and source properties (Figures~\ref{fig:clumpy_fc}--\ref{fig:diffinput_highNHI}). In clumpy media, the spectra approach the smooth-medium result as the covering factor increases, while at low \NHI, even a small number of interactions with rotating clumps can imprint the rotational Doppler shift. Halo spectra are also largely insensitive to the intrinsic \lya line width, indicating that they are primarily shaped by scattering in the rotating medium rather than by the source spectrum.
\end{enumerate}

These results suggest that spatially resolved \lya spectroscopy can provide an important diagnostic of CGM rotation, especially when combined with galaxy orientation, disk kinematics, and complementary absorption-line constraints. In particular, comparing spectra from opposite sides of the \lya halo may help distinguish rotational Doppler shifts from frequency diffusion in high-optical-depth gas. Future work should extend the present idealized models by including more realistic density distributions, dust, radial inflows/outflows, and observational effects such as finite spectral and spatial resolution, enabling direct comparison with IFU observations of extended \lya halos.

\section*{Acknowledgements}

H.-G. Kym and K.-I. Seon were supported by the Korea Astronomy and Space Science Institute under the R\&D program (Project No. 2026-1-831-03) supervised by the Korea AeroSpace Administration. K.-I. Seon was also supported by a National Research Foundation grant of Korea (NRF; No.\ 2020R1A2C1005788) funded by the Korea government.
SJC acknowledges support from the ERC synergy grant 101166930 - RECAP. 
MG thanks the European Union for support through ERC-2024-STG 101165038 (ReMMU).

\section*{Data Availability}

The data underlying this article will be shared on reasonable request to the first author.


\bibliographystyle{mnras}
\bibliography{reference}



\appendix

\section{Velocity maps in rotating outflow models}
\label{sec:appendix_rotating outflow}

\begin{figure*}
    \includegraphics[width = 0.99 \textwidth]{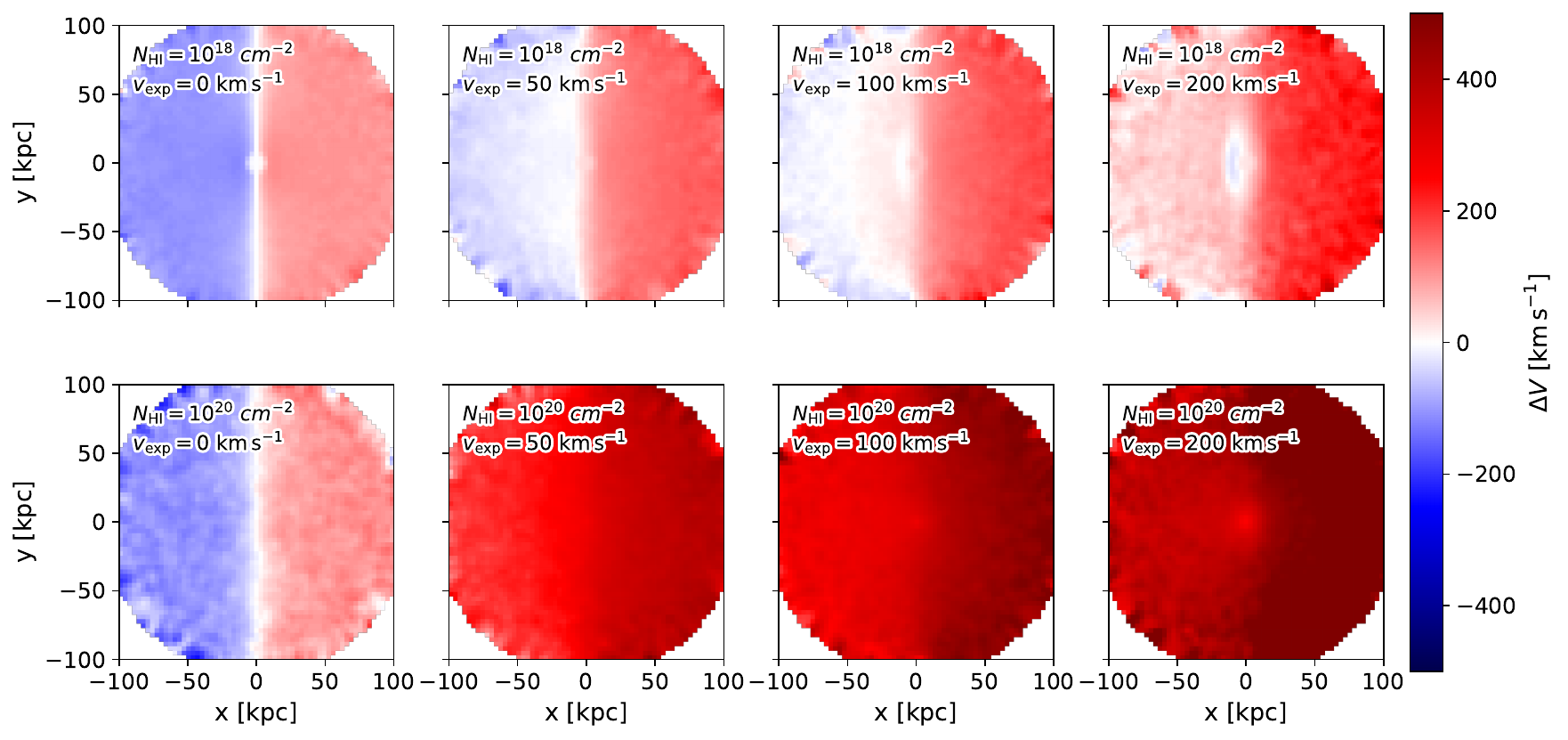}
    \caption{
    First-moment velocity maps for rotating outflow models with \vrot and \sigsrc fixed at 100 \kms. 
    The top and bottom rows show maps for \NHI = $10^{18}$ and $10^{20} \unitNHI$, respectively.
    From left to right, the radial expansion velocity \vexp increases from 0 (no outflow) to 200 \kms. 
    The maps are smoothed assuming MUSE-like spatial and spectral resolutions.
    At low \NHI, the red-blue pattern produced by rotation remains visible at $\vexp \le 100 \kms$.
    However, at high \NHI, the velocity map already shows predominantly redshifted features at $\vexp = 50 \kms$, indicating that multiple scatterings in the outflow medium cause significant redshift of escaping photons.
    }
    \label{fig:velocity_map_high_low_NHI}
\end{figure*}

In Section~\ref{sec:discussion_velocity_map}, we showed that a combination of rotation and radial outflow can produce complex first-moment velocity maps, and that such maps may not provide a direct representation of the intrinsic gas velocity field. Here, we further examine how this behavior depends on the H~I column density \NHI by comparing rotating outflow models with two different values of \NHI. This test is intended to illustrate the role of \lya radiative-transfer effects in shaping IFU-like velocity maps, rather than to provide a full exploration of radial outflow models.

Figure~\ref{fig:velocity_map_high_low_NHI} shows first-moment velocity maps for models with fixed \vrot at 100 \kms, while \vexp is varied from 0 to 200~\kms. 
In the top panels for $\NHI = 10^{18} \unitNHI$, the red-blue pattern produced by rotation remains visible for moderate outflow velocities, although the redshifted side becomes progressively stronger as \vexp increases. In contrast, in the bottom panels for $\NHI = 10^{20} \unitNHI$, even a relatively small radial expansion velocity ($\vexp = 50 \kms)$ produces a predominantly redshifted velocity field across most of the halo. This occurs because photons experience multiple redward frequency shifts through scattering in a higher \NHI medium, allowing the radial velocity field to imprint a stronger redshift on the emergent \lya photons. Therefore, the apparent dominance of outflow-like signatures in \lya velocity maps can be amplified by optical-depth effects, reinforcing the need to interpret spatially resolved \lya kinematics with radiative-transfer modeling.

\section{Degeneracy between the static and rotating models}
\label{sec:appendix_nhi_degeneracy}

\begin{figure*}
    \includegraphics[width=0.48\textwidth]{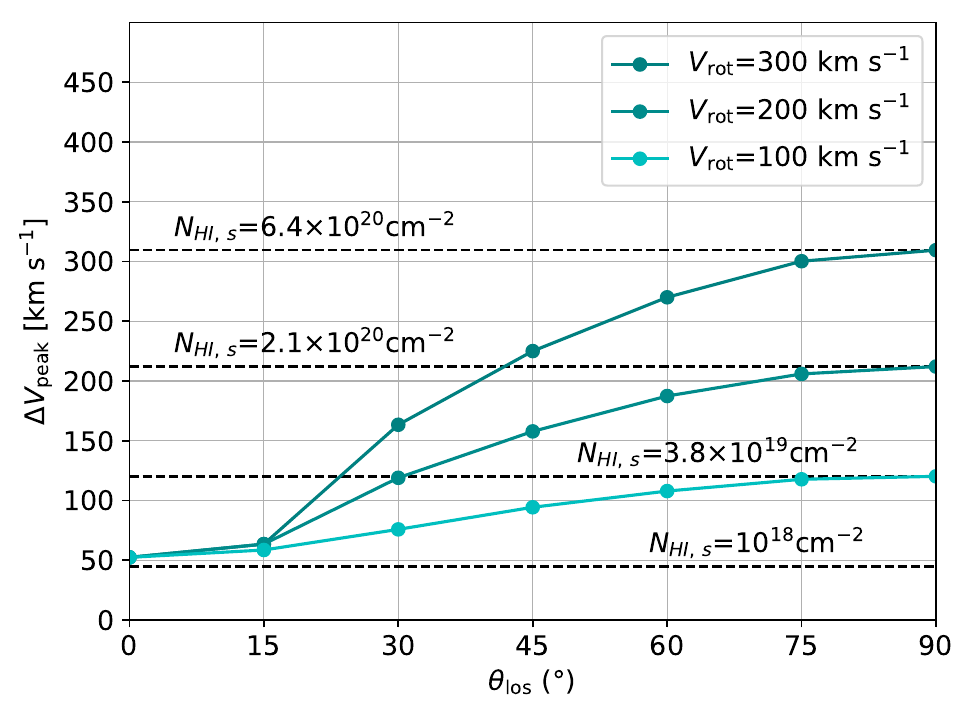}
    \includegraphics[width=0.48\textwidth]{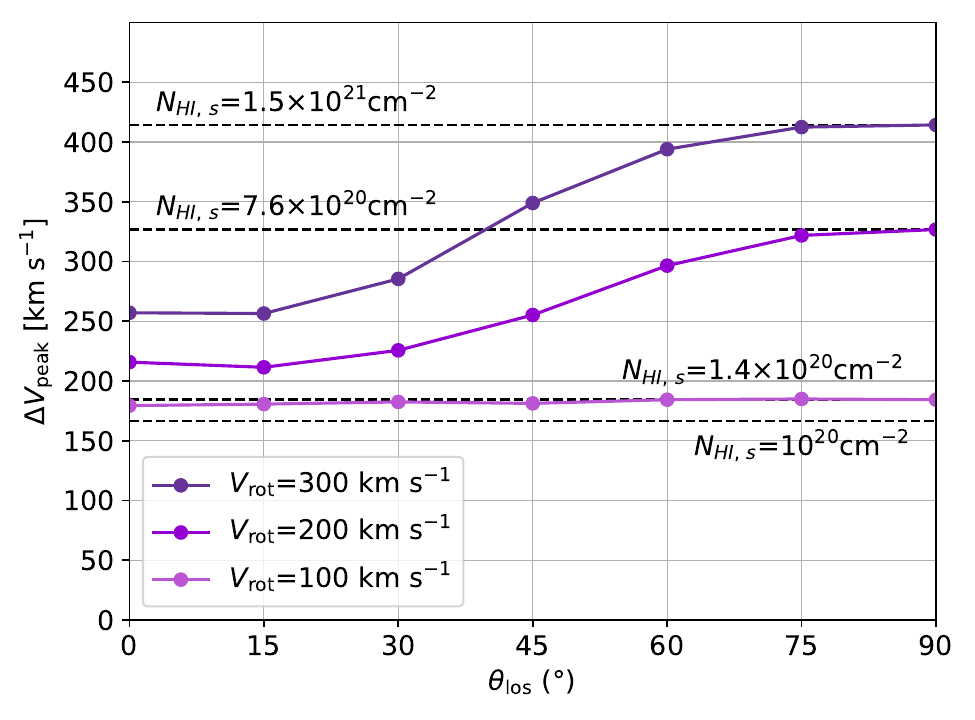}
    \caption{Peak positions (\dvpeak) as a function of viewing angle (0\textdegree -- 90\textdegree).
    The column densities of the rotating cases are \NHI=$10^{18} \unitNHI$ (left panel)  and $10^{20} \unitNHI$ (right panel), respectively.
    Solid lines represent the peak positions of the rotating cases, while the black dashed lines show the \dvpeak values of the static cases that match those of the rotating cases at \anglelos=90\textdegree. The corresponding column densities of the static cases ($N_{\rm HI,s}$) are noted in the figure.
    }
    \label{fig:peaksep_los&vrot}
\end{figure*}

Figure~\ref{fig:peaksep_los&vrot} illustrates the degeneracy between the static and rotating cases discussed above. It shows the peak position, \dvpeak, as a function of viewing angle, \anglelos, for three rotational velocities ($\vrot=100, 200, 300\kms$). The underlying rotating medium is fixed at $\NHI = 10^{18}\unitNHI$ in the left panel and $\NHI = 10^{20}\unitNHI$ in the right panel. The dashed horizontal lines indicate the \dvpeak\ values produced by static, non-rotating models with the column densities $N_{\rm HI,s}$ shown in each panel. For each rotating curve, $N_{\rm HI,s}$ is the column density of the static model that reproduces the same \dvpeak as the rotating model at edge-on viewing angle ($\anglelos=90\degree$).

In the left panel ($\NHI = 10^{18}\unitNHI$), \dvpeak\ increases smoothly with \anglelos\ for all three rotational velocities and reaches its maximum in the edge-on view. This behavior is consistent with the rotational Doppler shift being directly imprinted on the emergent spectrum when frequency diffusion is weak. The static models required to reproduce the edge-on peak positions have $N_{\rm HI,s} \approx 3.8\times10^{19}, 2.1\times10^{20}$, and $6.4\times10^{20}\unitNHI$ for $\vrot=100, 200$, and $300\kms$, respectively. These values are two to three orders of magnitude larger than the actual column density of the rotating medium. In the face-on view ($\anglelos=0\degree$), where the line-of-sight projection of the rotational velocity vanishes, the three rotating curves converge to the $N_{\rm HI,s}=10^{18}\unitNHI$ baseline. This confirms that the residual peak separation in this limit arises purely from radiative transfer in the underlying medium.

The right panel ($\NHI = 10^{20}\unitNHI$) shows qualitatively different behavior. Even in the face-on view, \dvpeak\ is already $\sim$180--250\kms, because frequency diffusion through the optically thick medium produces a substantial peak shift. The dependence on \anglelos\ becomes progressively weaker as \vrot\ decreases: for $\vrot=100\kms$, the curve is nearly flat over the full range of viewing angles, and even for $\vrot=200\kms$, the angular variation is much smaller than in the low-\NHI\ case. When scattering already drives strong frequency diffusion, the additional contribution from rotational Doppler shifts has a relatively weak effect on the peak separation. The static counterparts of the edge-on rotating cases require $N_{\rm HI,s}$ between $1.4\times10^{20}$ and $1.5\times10^{21}\unitNHI$, again significantly larger than the actual column density of the rotating medium.

Taken together, the two panels quantify this degeneracy: the integrated peak position alone cannot distinguish a moderately rotating, low-column-density medium from a static, high-column-density one. The viewing-angle dependence can partially break this degeneracy in the optically thin regime, where rotation produces a strong angular trend while a static medium would show no \anglelos\ dependence. At high \NHI, however, the angular trend itself is suppressed by frequency diffusion. This reinforces the conclusion that integrated spectra must be combined with spatially resolved diagnostics, such as the halo and core spectra shown in Figure~\ref{fig:resolved_spec}, to break the degeneracy between rotation and column density effects.

\bsp	
\label{lastpage}
\end{document}